\documentclass[conference,a4paper]{IEEEtran}
\usepackage{amssymb}
\usepackage{subfigure}
\usepackage{amsmath}
\usepackage{amsfonts}
\usepackage{mathrsfs}
\usepackage{engord}
\usepackage[dvips]{graphicx}
\usepackage{threeparttable}
\usepackage{booktabs}
\usepackage{epsfig}
\usepackage{color}
\usepackage{graphicx}
\usepackage{multirow}
\usepackage{cite}
\usepackage{epstopdf}
\usepackage{framed}
\usepackage{url}
\usepackage{pmat}
\usepackage{amsmath,bm}
\newtheorem{prop}{Proposition}

\newtheorem{defn}{Definition}

\begin{document}
\title{Topological and Algebraic Properties of Chernoff Information between Gaussian Graphs}
\author{\IEEEauthorblockN{Binglin Li, Shuangqing Wei, Yue Wang, Jian Yuan}}
\maketitle
\let\thefootnote\relax\footnotetext{B. Li, Y. Wang and J. Yuan are with Department of Electronic Engineering, Tsinghua University,
Beijing, P. R. China, 100084. (E-mail: {libl13}@mails.tsinghua.edu.cn; {wangyue, jyuan}@mail.tsinghua.edu.cn). S. Wei is
with the school of Electrical Engineering and Computer Science, Louisiana State University, Baton Rouge, LA 70803,
USA (Email: swei@lsu.edu). This material is based upon work supported in part by the National Science
Foundation (USA) under Grant No. 1320351, and the National Natural Science Foundation of China under Grant 61673237. }

\begin{abstract}

In this paper, we want to find out the determining factors of Chernoff information in distinguishing a set of Gaussian graphs. We find that Chernoff information of two Gaussian graphs can be determined by the generalized eigenvalues of their covariance matrices. We find that the unit generalized eigenvalues do not affect Chernoff information and their corresponding dimensions do not provide information for classification purpose.
In addition, we can provide a partial ordering using Chernoff information between a series of Gaussian trees connected by independent grafting operations.
With the relationship between generalized eigenvalues and Chernoff information, we can do optimal classification linear dimension reduction with least loss of information for classification.

\end{abstract}

\section{Introduction}

Gaussian graphical models are widely used in constructing the conditional independence of continuous random variables. It is used in many applications such as social networks\cite{vega2007complex}, economics\cite{dobra2010modeling}, biology\cite{ahmed2008time} and so on.  Among Gaussian graphical models, we are particularly interested in Gaussian
trees because of its sparse structure and the existence of computationally
efficient algorithms in learning the underlying topologies. 

{In our study, we focus on classification i.e. hypothesis testing against a set of given Gaussian distributions with sparse graph structures.  In this $M$-ary hypothesis testing problem, 
we infer which hypothesis a data sequence is generated from. 
The error probability decreases while data sequence size increases. So we use error exponent to measure how fast error probability decreases along with data. 
Error exponent is important when we want to
estimate how much testing data we need to achieve a given error probability.}

%%The error events in {hypothesis testing problem} have two different types, namely, conditional error events and average error events. Conditional error events mean the error events between true model and the learned one when we learn an  approximate model from the output data\cite{Tan.TSP.Ana.2010,tan2011large,jog2015model}. This kind of errors also appear in network\cite{qiao1995,qiao2000}. We can use Kullback-Leibler (KL) distance to quantify this error exponent on distinguishing true model from the learnt one\cite{kullback1951information}.%%

In particular, we aim at the error exponent associated with average error probabilities. The resulting error exponent characterizing the vanishing rate of average error probability approaching zero is thus determined by the minimum Chernoff information among all  M-choose-2 pairs of hypotheses \cite{cover2012elements}. It should be noted that in literature (e.g.\cite{chernoff1952measure,westover2008asymptotic}),  because of the complexity in  attaining closed form solutions to Chernoff information, KL distance was often adopted as a bound to Chernoff information.

In algebraic analysis of hypothesis testing problem, we also use generalized eigenvalues of covariance matrices as a metric of the difference between them\cite{5540052,6247965}.
Clearly, Chernoff information and generalized eigenvalues of covariance matrices are respectively probabilistic and algebraic ways to describe the difference between two Gaussian graphs. There must be  relation among topology, statistical distributions (Chernoff information), and algebra (generalized eigenvalues). This paper shows how Chernoff information can be determined by generalized eigenvalues. In addition, we show how topological differences affect generalized eigenvalues and thus Chernoff information. Our work, to the best of our knowledge, is the first one investigating such relationship from Chernoff information point of view.

More specifically, we find that two Gaussian graphs can be linearly and inversely transformed to two graphs whose covariance matrices are diagonal. Entries of the diagonal matrices are related to generalized eigenvalues. Thus we find that Chernoff information between two Gaussian graphs is an expression of generalized eigenvalues and a special parameter $\lambda^*$, which is also determined by generalized eigenvalues. In addition, we find that the unit generalized eigenvalues do not affect Chernoff information and the corresponding dimensions make no contribution to differentiating two Gaussian graphs for classification problem.

Our former paper \cite{binglin} dealt with the classification problem related to Gaussian trees. We found that some special operations on Gaussian trees, namely adding operation and division operation, do not change Chernoff information between them. Now in this paper, we find that these two operations only add one extra unit generalized eigenvalue and do not affect other generalized eigenvalues. We can
use generalized eigenvalues to prove the same proposition.
Paper\cite{binglin} also dealt with two Gaussian trees connected by one grafting operation and showed that Chernoff information between them is the same with that of two special $3$-node trees whose weights are related to the underlying operation. In this paper, we extend this result to a Gaussian tree chain connected by independent grafting operations and provide a partial ordering of Chernoff information between these trees.

In practical scenarios, we may not have access to all the output of the model. Instead, we may have some constraint on observation costs, which prompts us to reduce the dimension of observation vectors  in order to meet such  constraints\cite{NOWAKOWSKA201674,GUAN2013147}. A good choice here is doing linear dimension reduction in collection stage. {We name this dimension reduction as classification dimension reduction to distinguish it from traditional dimension reduction.} We only deal with a $2$-ary hypothesis testing in this part. {We linearly transform an $N$ dimensional Gaussian vector $\mathbf{x}$ to an $N_O < N$ dimensional vector $\mathbf{y}=\mathbf{A}\mathbf{x}$, through an $N_O \times N$ matrix $\mathbf{A}$. We want to find the optimal linear transformation $\mathbf{A}^*$ which can maximize Chernoff information {of two low-dimensional distributions}.} Our former paper \cite{binglin} only dealt with a simple, but non-trivial case with $N_O = 1$.
In this work, we  offer an optimal method to maximize the resulting Chernoff information after a linear transformation for an arbitrary $N_O \geq 1$.

{We can divide the features of two distributions into two parts, namely shared features and discrepant features. The aim of classification dimension reduction is to keep discrepant features while discard shared features.}
Traditional dimension reduction methods, such as Principal Component Analysis (PCA) and other Representation Learning\cite{6472238}, aim to find the optimal features with maximum information.
{In traditional dimension reduction, we can also divide the features of high-dimensional distributions into two parts. But they are main features and minor features. The aim of traditional dimension reduction is to keep main features while discard minor features. In this way,  traditional dimension reduction can recover most high-dimensional informtion from low-dimensional data. Our classification dimension reduction problem have different purpose compared to traditional dimension reduction methods. Some important features in traditional dimension reduction methods may be useless in our method because these features in two hypotheses are similar. In addition, our method needs to compare two distributions, while traditional dimension reduction methods, however, only consider one distribution.}

Our major and novel results can be summarized as follows. We first provide the relationship between Chernoff information and generalized eigenvalues, which shows that generalized eigenvalues which are equal to $1$ make no contribution to Chernoff information. We use this result to explain why adding and division operations of \cite{binglin} do not affect Chernoff information between Gaussian trees. These results build a relationship between topology, statistical distribution and algebra.
In addition, we deal with Gaussian trees connected by more than one grafting operation and show a partial ordering inside the chain. At last, we provide an optimal classification linear dimension reduction method.

This paper is organized as follows. In Section \ref{s2}, we propose the models of our analysis. The relationship between topology, Chernoff information and generalized eigenvalues is shown in Section \ref{s3}. The partial ordering of Chernoff information in independent grafting chain is presented in  Section \ref{s4}.  Section \ref{s5} shows the optimal classification linear dimension reduction method. In Section \ref{s6}, we conclude the paper.

\section{System Model}\label{s2}

Gaussian tree models can represent the dependence of multiple Gaussian random variables by tree topologies. For simplification, we normalize the variance of all Gaussian variables to be $1$ and the mean values to be $0$. For an $N$-node tree $\mathbf{G} = (V, E, W)$ with  vertex set $V=\{1,\dots,N\}$, edge set $E=\big\{e_{ij}|(i,j)\subset V\times
V\big\}$ and edge weights set $W=\{w_{ij}\in [-1,1]|e_{ij}\in E\}$, $E$ satisfies $|E|=N-1$ and contains no cycles. A distribution $\mathbf{x}=[x_1,x_2,\dots,x_N]^T\sim N(\mathbf{0},\Sigma)$ is said to be a normalized Gaussian distribution on the tree $\mathbf{G} = (V, E, W)$ if
    \begin{align}
    \sigma_{ij}=
    \begin{cases}
    1&\quad i=j\\
    w_{ij}&\quad e_{ij}\in E\\
    w_{im}w_{mn}\dots w_{pj}&\quad e_{ij}\notin E
    \end{cases}
    \end{align}
    where $\sigma_{ij}$ is the $(i,j)$ term of $\Sigma$ and $e_{im}e_{mn}\dots e_{pj}$ is the unique path from node $i$ to node $j$.

    A normalized covariance matrix of a Gaussian tree has a very simple inverse matrix and determinant, as shown in Proposition \ref{thm1} which has been proved  in our former paper\cite{binglin}.

    \begin{prop}\label{thm1}
    Assume $\Sigma$ is a normalized covariance matrix of Gaussian tree  $G=(E,V,W)$, so $|\Sigma|=\prod_{e_{ij}\in E}(1-w_{ij}^2)$ and the elements $[u_{ij}]$ of $\Sigma^{-1}$ follow the following expressions:
    \begin{align}
    u_{ij}=
    \begin{cases}
    \frac{-w_{ij}}{1-w_{ij}^2}&\quad i\neq j~\text{and}~e_{ij}\in E\\
    0&\quad i\neq j~\text{and}~e_{ij}\notin E\\
    1+\sum_{p:e_{ip}\in E}\frac{w_{ip}^2}{1-w_{ip}^2}& \quad i=j.
    \end{cases}
    \end{align}
    \end{prop}

Consider a set of Gaussian trees, namely, $G_k(\mathbf{x}), k=1,2,\dots,M$, with their prior probabilities given by $\pi_1,\pi_2,\dots,\pi_M$.
We want to do
an $M$-ary hypothesis testing to find out from which Gaussian distribution
the data sequence $\mathbf{X}=[\mathbf{x}_1,\dots,\mathbf{x}_t]$
($\mathbf{x}_l={[x_{1,l},\dots,x_{N,l}]}^T$) comes from. We define the average error
probability of the hypothesis testing to be $P_e$, and let $E_e = \lim_{ t\rightarrow\infty}\frac{-\ln P_e}{t}$ be
the resulting error exponent,  which
depends on the smallest Chernoff information between the
trees \cite{westover2008asymptotic}
, namely
    \begin{align}
    E_e=\min_{1\leq i\neq j\leq M} CI(\mathbf{\Sigma}_i||\mathbf{\Sigma}_j)
    \end{align}
    where $CI(\mathbf{\Sigma}_i||\mathbf{\Sigma}_j)$ is the Chernoff information between the $i^{th}$ and $j^{th}$ trees.

For two $\mathbf{0}$-mean N-dim Gaussian joint distributions, $\mathbf{x}_1\sim N(0,\mathbf{\Sigma}_1)$ and $\mathbf{x}_2\sim N(0,\mathbf{\Sigma}_2)$, their Kullback-Leibler divergence is as follows
    \begin{align}
    D(\mathbf{\Sigma}_1||\mathbf{\Sigma}_2)=
    \frac{1}{2}\ln\frac{|\mathbf{\Sigma}_2|}{|\mathbf{\Sigma}_1|}+\frac{1}{2}tr(\mathbf{\Sigma}_2^{-1}\mathbf{\Sigma}_1)-\frac{N}{2}\label{D}
    \end{align}
    where $tr(\mathbf{X})=\sum_{i}x_{ii}$ is the trace of square matrix $\mathbf{X}$.
    We define a new distribution $N(0,\mathbf{\Sigma}_\lambda)$ in the exponential family of the $N(0,\mathbf{\Sigma}_1)$ and $N(0,\mathbf{\Sigma}_2)$, namely
    \begin{align}
    \mathbf{\Sigma}_\lambda^{-1}=\mathbf{\Sigma}_1^{-1}\lambda+\mathbf{\Sigma}_2^{-1}(1-\lambda)
    \end{align}
so that Chernoff information can be given as
    \begin{align}
    CI(\Sigma_1||\Sigma_2)=D(\Sigma_{\lambda^*}||\Sigma_2)=D(\Sigma_{\lambda^*}||\Sigma_1)\label{lambda}
    \end{align}
    where $\lambda^*$ is the unique point in $[0,1]$ at which  the latter equation is satisfied\cite{cover2012elements}.

    We already know that the overall Chernoff information in an $M$-ary testing is bottlenecked by the minimum pair-wise difference\cite{cover2012elements}, thus we next focus on the calculation of Chernoff information of  pair-wise Gaussian trees.

{ We also consider classification dimension reduction problem in 2-ary hypothesis testing problem.
 If we can only observe a low $N_O$-dim vector, namely $\mathbf{y}=\mathbf{A}\mathbf{x}$, where $\mathbf{A}$ is an $N_O\times
    N$ matrix and $\mathbf{x}\in R^N,\mathbf{y}\in R^{N_O}$, the new low dimensional
    variables follow joint distributions $N(\mathbf{0},\hat{\Sigma}_1)$
    and  $N(\mathbf{0},\hat{\Sigma}_2)$, where $\hat{\mathbf{\Sigma}}_i=\mathbf{A}\mathbf{\Sigma}_i\mathbf{A}^T, i=1,2$.  For fixed $N_O$, we want to
    find out the optimal $\mathbf{A}^*$ and its Chernoff information result
    $CI(\hat{\mathbf{\Sigma}}_1^*||\hat{\mathbf{\Sigma}}_2^*)$, s.t.
    \begin{align}
    \mathbf{A}^*=\arg \max_{\mathbf{A}} CI(\hat{\mathbf{\Sigma}}_1||\hat{\mathbf{\Sigma}}_2)\label{2A}
    \end{align}

\section{Generalized eigenvalues, Chernoff information and topology}\label{s3}

        Chernoff information is the measurement of the difference between statistical distributions. It is hard to be calculated directly and we rarely study its insights about the relationship between Chernoff information and structure characters. Chernoff information and generalized eigenvalues are both important parameters to describe the difference of Gaussian distributions. We expose the relationship among generalized eigenvalues, Chernoff information and topology.

\subsection{Linear transformation to diagonal covariance matrix related to generalized eigenvalues}\label{3A}

For two $N$-node $0$-mean Gaussian graphs $G_1$ and $G_2$ on random variables $\mathbf{x}$, whose covariance matrices are $\mathbf{\Sigma}_1$ and $\mathbf{\Sigma}_2$, we can use an inverse linear transformation matrix $\mathbf{P}$ to transform them to $\mathbf{x}'=\mathbf{P}\mathbf{x}$ whose covariance matrices $\mathbf{\Sigma}_1'$ and $\mathbf{\Sigma}_2'$ are diagonal and related to the generalized eigenvalues of $\mathbf{\Sigma}_1$ and $\mathbf{\Sigma}_2$.

$\mathbf{\Sigma}_1$ and $\mathbf{\Sigma}_2$ are real symmetric positive definite matrices, so the eigenvalues of $\mathbf{\Sigma}_1\mathbf{\Sigma}_2^{-1}$ are all positive, as shown in Appendix \ref{A1}.
The eigenvalue decomposition of $\mathbf{\Sigma}_1\mathbf{\Sigma}_2^{-1}$ is $\mathbf{Q}\mathbf{\Lambda} \mathbf{Q}^{-1}$, where $\mathbf{Q}$ is an $N\times N$ matrix and $\mathbf{\Lambda}=Diag(\{\lambda_i\})$ is a diagonal matrix of eigenvalues, in which we put multiple eigenvalues adjacent. $\{\lambda_i\}$ are the eigenvalues of $\mathbf{\Sigma}_1\mathbf{\Sigma}_2^{-1}$, namely the generalized eigenvalues of $\mathbf{\Sigma}_1$ and $\mathbf{\Sigma}_2$.
Note that $\mathbf{Q}$ may be non-orthogonal when $\mathbf{\Sigma}_1\mathbf{\Sigma}_2^{-1}$ is not symmetric.

\begin{prop}\label{prop0}
For two $N$-node $0$-mean Gaussian graphs $G_1$ and $G_2$ whose covariance matrices are $\mathbf{\Sigma}_1$ and $\mathbf{\Sigma}_2$ respectively,
we can construct a linear transformation matrix $\mathbf{P}={\left(\mathbf{Q}^{-1}\mathbf{\Sigma}_2{(\mathbf{Q}^{-1})}^T\right)}^{-\frac{1}{2}}\mathbf{Q}^{-1}$ and thus
\begin{align}
\mathbf{\Sigma}_2'=&\mathbf{P}\mathbf{\Sigma}_2\mathbf{P}^T=\mathbf{I}_N\\
\mathbf{\Sigma}_1'=&\mathbf{P}\mathbf{\Sigma}_1\mathbf{P}^T=\mathbf{\Lambda}
\end{align}
where eigenvalue decomposition of $\mathbf{\Sigma}_1\mathbf{\Sigma}_2^{-1}$ is $\mathbf{Q}\mathbf{\Lambda} \mathbf{Q}^{-1}$.
\end{prop}

The proof of Proposition \ref{prop0} is shown in Appendix \ref{A1}.

 We can treat $\mathbf{\Sigma}_1'$ and $\mathbf{\Sigma}_2'$ as two covariance matrices of graphs $G_1'$ and $G_2'$ on $\mathbf{x}'$ obtained from $G_1$ and $G_2$ by inverse linear transformation $\mathbf{P}$.
$G_1'$ and $G_2'$ are graphs with $N$ independent variables.

The distances of $G_1'$ and $G_2'$ are as follows. The distances between $G_1$ and $G_2$ are the same with the distances of $G_1'$ and $G_2'$  because $\mathbf{P}$ is inverse.

\begin{align}
D(\mathbf{\Sigma}_1||\mathbf{\Sigma}_2)=&D(\mathbf{\Sigma}_1'||\mathbf{\Sigma}_2')=\frac{1}{2}\sum_i \left( -\ln\lambda_i+\lambda_i-1\right)\\
D(\mathbf{\Sigma}_2||\mathbf{\Sigma}_1)=&D(\mathbf{\Sigma}_2'||\mathbf{\Sigma}_1')=\frac{1}{2}\sum_i \left( \ln\lambda_i+\frac{1}{\lambda_i}-1\right)\\
D(\mathbf{\Sigma}_\lambda||\mathbf{\Sigma}_1)=&D(\mathbf{\Sigma}_\lambda'||\mathbf{\Sigma}_1')\nonumber\\=\frac{1}{2}\sum_i&\left(\ln\left(\lambda+(1-\lambda)\lambda_i\right)+\frac{1}{\lambda+(1-\lambda)\lambda_i}-1\right)\\
D(\mathbf{\Sigma}_\lambda||\mathbf{\Sigma}_2)=&D(\mathbf{\Sigma}_\lambda'||\mathbf{\Sigma}_2')\nonumber\\=\frac{1}{2}\sum_i&\left(\ln\frac{\lambda+(1-\lambda)\lambda_i}{\lambda_i}+\frac{\lambda_i}{\lambda+(1-\lambda)\lambda_i}-1\right)\\
CI(\mathbf{\Sigma}_1||\mathbf{\Sigma}_2)=&CI(\mathbf{\Sigma}_1'||\mathbf{\Sigma}_2')=D(\mathbf{\Sigma}_{\lambda^*}'||\mathbf{\Sigma}_1')=D(\mathbf{\Sigma}_{\lambda^*}'||\mathbf{\Sigma}_2')\label{CI}
\end{align}
where $\mathbf{\Sigma}_\lambda^{-1}=\mathbf{\Sigma}_1^{-1}\lambda+\mathbf{\Sigma}_2^{-1}(1-\lambda)$ and ${\mathbf{\Sigma}_\lambda'}^{-1}={\mathbf{\Sigma}_1'}^{-1}\lambda+{\mathbf{\Sigma}_2'}^{-1}(1-\lambda)$. Matrix $\mathbf{\Sigma}_\lambda'$ is also  diagonal. Parameter
$\lambda^*$ in (\ref{CI}) is the unique root of $D(\mathbf{\Sigma}_{\mathbf{\lambda}^*}'||\mathbf{\Sigma}_1')=D(\Sigma_{\mathbf{\lambda}^*}'||\mathbf{\Sigma}_2')$, namely
$\sum_i\left(\frac{1-\lambda_i}{\lambda^*+(1-\lambda^*)\lambda_i}+\ln\lambda_i\right)=0$.

In this way, we can conclude that the KL and CI divergences between two Gaussian graphs can be determined by their generalized eigenvalues.

\subsection{Relationship between generalized eigenvalues and Chernoff information}

        Here we show the relationship between Chernoff information of two Gaussian graphs and  generalized eigenvalues of their covariance matrices $\mathbf{\Sigma}_1$ and $\mathbf{\Sigma}_2$. We define $\prod_{i=1}^{N} \lambda_i=|\mathbf{\Sigma}_1|/|\mathbf{\Sigma}_2|=\beta$ here.

    \begin{prop}\label{prop1}
 For two $N$-node Gaussian distributions whose covariance matrices are $\mathbf{\Sigma}_1,\mathbf{\Sigma}_2$ and $|\mathbf{\Sigma}_1|/|\mathbf{\Sigma}_2|=\beta$, their Chernoff information satisfies
    \begin{align}
    &CI(\mathbf{\Sigma}_1||\mathbf{\Sigma}_2)=\nonumber\\
    &\frac{1}{2} \sum_i \left\{\ln\{(1-\lambda^*) \sqrt{\lambda_i}+\frac{\lambda^*}{\sqrt{\lambda_i}}\}\right\}+ \frac{1}{2}(\lambda^*- \frac{1}{2})\ln\beta
    \end{align}
    where $\{\lambda_i\}$ are the generalized eigenvalues of $\mathbf{\Sigma}_1,\mathbf{\Sigma}_2$,  and
    $\lambda^*\in[0,1]$ is the unique result of
    \begin{align}
\sum_{i} \frac{1}{\lambda^*+(1-\lambda^*)\lambda_i}=N+(\lambda^*-1)\ln\beta\label{2}\\
\sum_{i} \frac{\lambda_i}{\lambda^*+(1-\lambda^*)\lambda_i}=N+\lambda^*\ln\beta\label{1}
    \end{align}
    \end{prop}

    The results of equation (\ref{2}) and (\ref{1}) are the same. We can prove this proposition from  equation (\ref{CI}), as shown in Appendix \ref{A2}.

    We  find that  generalized eigenvalues of covariance matrices $\mathbf{\Sigma}_1$ and $\mathbf{\Sigma}_2$ are the key parameters of Chernoff information. We can get  Chernoff information with these $N$ generalized eigenvalues, so these $N$ parameters contain all the information about the difference between two Gaussian trees.
    The generalized eigenvalues are so important that we need more properties about them.

\subsection{Performance of unit generalized eigenvalues}

In former equation, we find that unit generalized eigenvalues are very special to Chernoff information.

    \begin{prop}\label{prop3}
    Assuming that the generalized eigenvalues of $(N+1)$-node $G_{1}'$ and $G_{2}'$ are the same with that of $N$-node $G_{1}$ and $G_{2}$ except a newly additional unit generalized eigenvalue, the optimal parameter $\lambda^*$ of $(\mathbf{\Sigma}_1',\mathbf{\Sigma}_2')$ is the same with that of $(\mathbf{\Sigma}_1,\mathbf{\Sigma}_2)$ and $CI(\mathbf{\Sigma}_1'||\mathbf{\Sigma}_2')= CI(\mathbf{\Sigma}_1||\mathbf{\Sigma}_2)$.
    \end{prop}

Under this assumption, $\frac{|\mathbf{\Sigma}_1'|}{|\mathbf{\Sigma}_2'|} =(\prod_{i=1}^N\lambda_i)\times 1=\beta=\frac{|\mathbf{\Sigma}_1|}{|\mathbf{\Sigma}_2|} $.
Proposition \ref{prop3} can be proved from Proposition \ref{prop1} as shown in Appendix \ref{A3}.

    Proposition \ref{prop3} shows a possible way to do dimension reduction that we can reduce the dimension of Gaussian graphs from $N+1$ to $N$ without decreasing their Chernoff information.

\begin{figure}
    \centering
    \includegraphics[width=4.5cm]{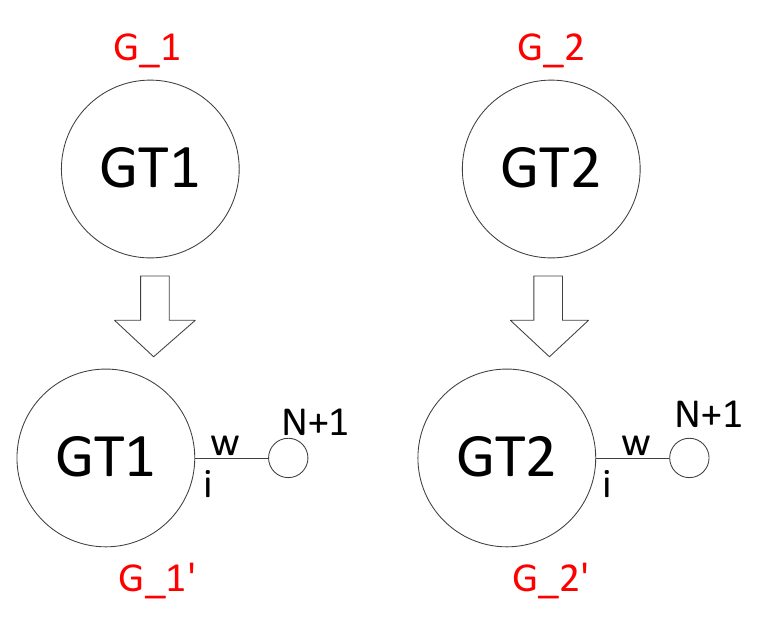}\\
    \includegraphics[width=7cm]{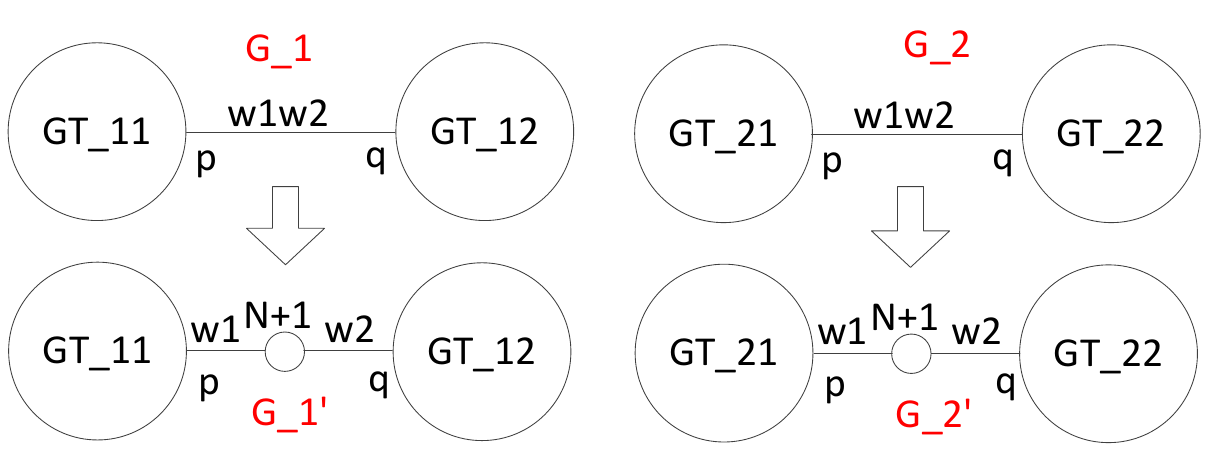}\\
    \caption{Adding and Division operations of two trees}\label{division}\label{adding}
    \end{figure}

Paper\cite{binglin} dealt with the classification on Gaussian trees. In that paper, we defined two   special operations on Gaussian tree pairs, namely adding operation and division operation as shown in Fig. \ref{adding}. The circles in the figure represent Gaussian trees. For adding operation, we add the same leaf node $N+1$, which has the same neighbor $i$ with weight $w$, to both trees. Division operation only appears when two trees have the same edge $e_{pq}$ with the same weight $w_1w_2$, for which we split this edge into two edges and add a node $N+1$ in the path of $p-q$ which has edges $e_{(N+1)p}$ and $e_{(N+1)q}$ with weights $w_1$ and $w_2$, respectively.
 {After these operations, we get a new pair of Gaussian trees with different dimensions compared to original Gaussian tree pairs.}
    These two operations do not change Chernoff information between two Gaussian trees. Next we show how generalized eigenvalues change after adding or division  operation. The covariance matrices of Gaussian trees are normalized.

    \begin{prop}\label{prop2}
    Assume that Gaussian trees $G_{1}'$ and $G_{2}'$ are obtained from $G_{1}$ and $G_{2}$ by adding operation or division operation. Their covariance matrices are $(\mathbf{\Sigma}_1',\mathbf{\Sigma}_2'), (\mathbf{\Sigma}_1,\mathbf{\Sigma}_2)$ respectively. The generalized eigenvalues of $(\mathbf{\Sigma}_1',\mathbf{\Sigma}_2')$ are the same with that of $(\mathbf{\Sigma}_1,\mathbf{\Sigma}_2)$ except a newly added unit eigenvalue.
    \end{prop}

        Proposition \ref{prop2} is proved in Appendix \ref{A4}. From Proposition \ref{prop3} and \ref{prop2},  we can conclude that adding and division operations do not affect Chernoff information between two Gaussian trees, which has been proved in our former paper \cite{binglin}. We prove it using generalized eigenvalues now.

\section{Partial ordering in independent grafting chain}\label{s4}

Grafting operation is a kind of topological operation by cutting down a subtree from another tree and pasting it to another location as shown in Fig.~\ref{grafting}. In this figure, $i,p,q$ are the nodes in both trees, representing random variables $x_i,x_p,x_q$. We separate subtree1 and subtree2 by cutting the edge $e_{ip}$ with weight $w$ and paste subtree2 to subtree1 by adding new edge $e_{iq}$ with the same weight $w$.
{We do not actually cut any edges from Gaussian trees, though the name of the operation suggests otherwise. We use it to describe the topological difference between two Gaussian trees.}

 \begin{figure}
        \centering
        \includegraphics[width=8cm]{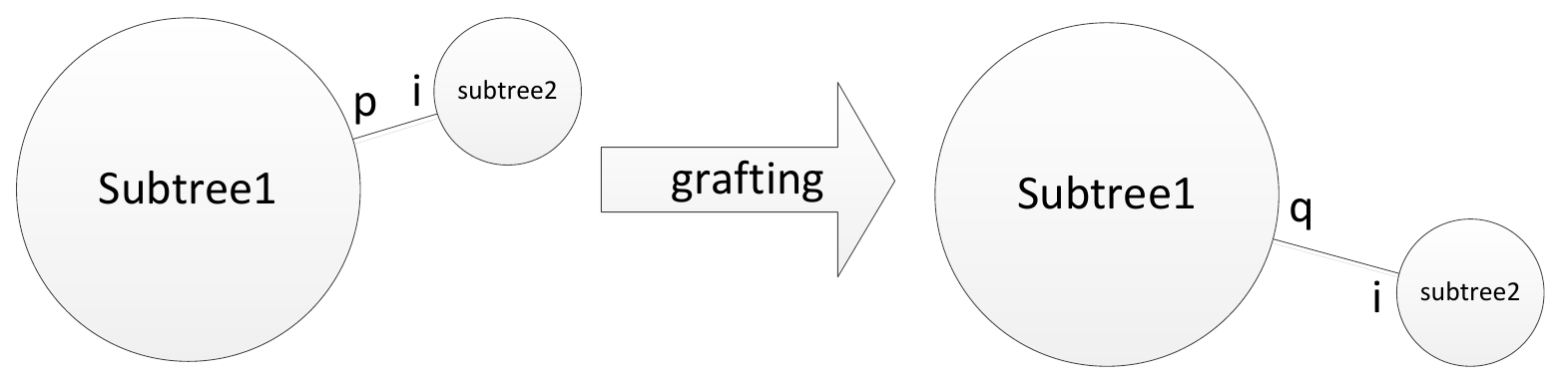}\\
        \caption{$T_2$ is obtained from $T_1$ by grafting operation}\label{grafting}
    \end{figure}

    In our former paper\cite{binglin}, we have shown that two Gaussian trees connected by one grafting operation have the same Chernoff information with two special $3$-node Gaussian trees whose weights are related to the underlying operation. Now we consider a more complex situation: two Gaussian trees connected by more than one grafting operation. According to Proposition \ref{thm1}, grafting operations do not change determinant of normalized Gaussian trees.

    Gaussian trees connected by more grafting operations can not be simplified to a fixed couple of small trees because the interaction of these grafting operations varies. Our initial expectation was that bigger difference in topology between two Gaussian trees leads to larger Chernoff information. This may not be true for all situations.

    Before we deal with a sequence of grafting operations, we need to constrain the interaction among them.  
    We  define the independence of grafting operations at first.

    \begin{figure}
    \centering
    \includegraphics[width=8cm]{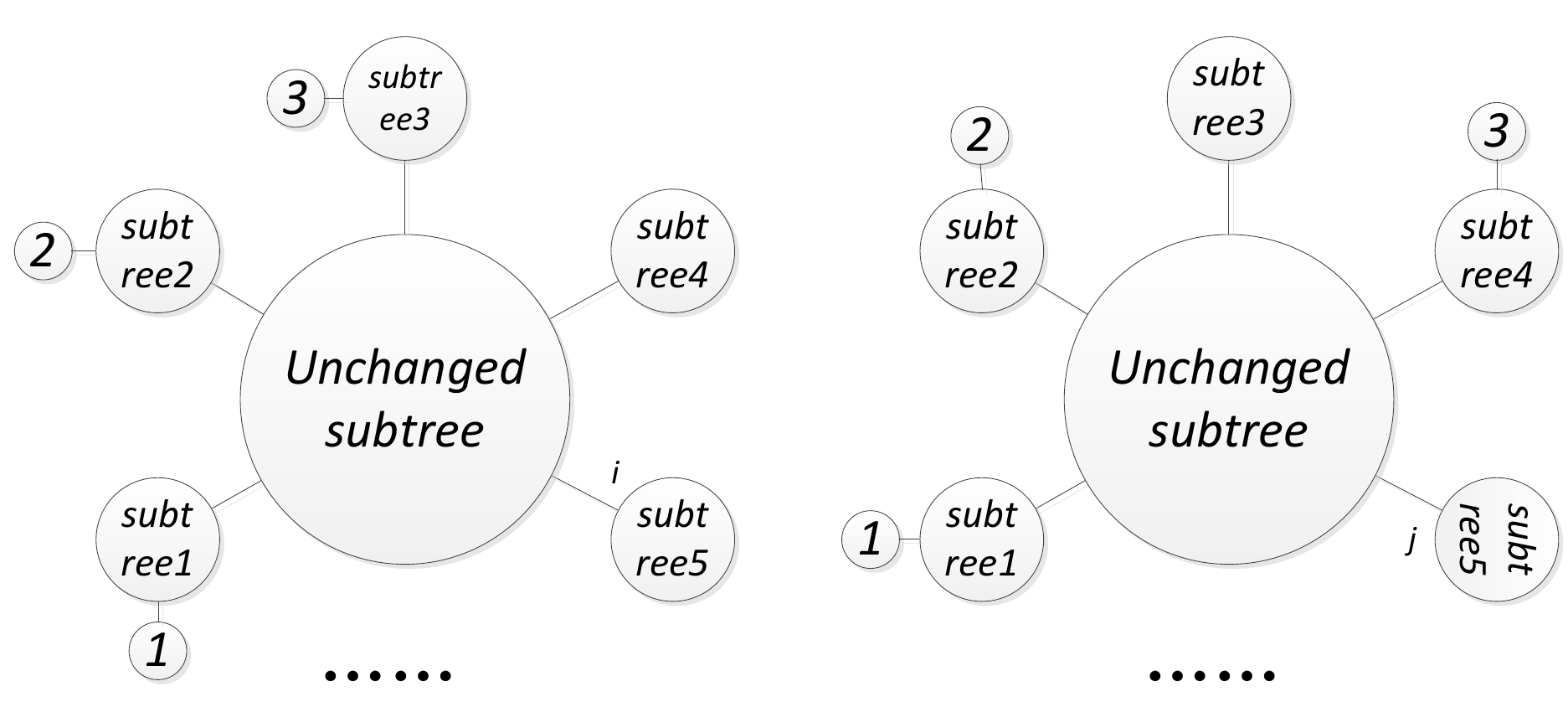}\\
    \caption{Independent grafting operations}\label{z}
    \end{figure}

    \begin{defn}
  If all the grafting operations can be divided into different subtrees, as shown in Fig. \ref{z}, then these grafting operations are independent. After regrouping all the nodes, the whole tree has star-shaped topology. The subtree in the center is unchanged during grafting operations. Grafting operations are involved in disjoint super leaf nodes of the star.
    \end{defn}

    In Fig. \ref{z}, we show $4$ independent grafting operations around the unchanged subtree. There are three types of grafting operations in the star-shaped topology. From left to right, the $1$-st, $2$-nd grafting operations belong to the first type, the $4$-th one belongs to the second type and the $3$-rd operation belongs to the third type. For the first type, we can cut a subtree, represented by a small circle with a number in it, from the super leaf node and paste it to another part of this super leaf node. In the second type, we can cut the unchanged subtree outside the super leaf node and paste it to another location in this super leaf node. But for the third type, we   cut a subtree, represented by a small circle with a number in it,   from a super leaf node and paste it to another super leaf node.  The third kind of grafting operations involve two super leaf nodes of the star while the first and second kinds of operations only involve one.

    If all the grafting operations are independent, then we can make the following conclusion.

    \begin{prop}\label{prop7}
    For two Gaussian trees connected by several independent grafting operations, $\lambda^*=1/2$ holds.
    \end{prop}

We can prove it as follows. 
    The trees have the same number of nodes and the same entropy due to grafting operations. Parameter $\lambda^*$ satisfies $tr(\mathbf{\Sigma}_{\lambda^*}(\mathbf{\Sigma}_1^{-1}-\mathbf{\Sigma}_2^{-1}))=0$, which can be transformed from the definition formulas of $\lambda^*$. Expression $tr(\mathbf{\Sigma}_{\lambda^*}(\mathbf{\Sigma}_1^{-1}-\mathbf{\Sigma}_2^{-1}))$ is a summation formula with $4n$ term, where each $4$ terms are related to one single grafting operation. We can deal with the terms respectively and prove $tr(\mathbf{\Sigma}_{0.5}(\mathbf{\Sigma}_1^{-1}-\mathbf{\Sigma}_2^{-1}))=0$ eventually.
    More details can be found in Appendix \ref{A5}.

    \begin{prop}\label{<}
    For the grafting chain $T_1\leftrightarrow T_2\leftrightarrow T_3 \leftrightarrow \dots \leftrightarrow T_n$ where all the grafting operations in the chain are independent, we can conclude that $CI(T_i||T_j)\leq CI(T_p||T_q)$ if $p\leq i\leq j\leq q$.
    \end{prop}

    Detail of the proof can be found in Appendix \ref{A6}.

        If we want to find out the minimum Chernoff information in this chain, we only need to try $n-1$ pairs of $T_i-T_{i+1}$, rather than all the  $\binom n 2$ pairs. The number of candidates is significantly reduced.

   We can not compare $CI(T_1||T_2)$ and $CI(T_2||T_3)$ even in a simple chain $T_1\leftrightarrow T_2 \leftrightarrow T_3$ without knowing the weights.
    We can only compare Chernoff information pairs $CI(T_i||T_j), CI(T_p||T_q)$ with $p\leq i\leq j\leq q$ ordering, so this result is a partial ordering inequality, rather than a full ordering inequality.

    In Proposition \ref{<}, we constrain the grafting operations independent. We may wonder whether the result suits for all the possible grafting chains   without independent assumption. Taking grafting chain $T_1\leftrightarrow T_2 \leftrightarrow T_3$ in Fig. \ref{12} as an example, the two grafting operations are not independent. Intuition tells us that $CI(T_1||T_3)$ is likely larger than $CI(T_1||T_2)$ and $CI(T_2||T_3)$, because the difference between $T_1$ and $T_3$ is the accumulation of $T_1-T_2$'s difference and   $T_2-T_3$'s difference.   Some numerical cases is shown in Table \ref{form2}. In this table, we can find that $CI(T_1||T_3)<CI(T_1||T_2)$ can hold. This is a counter-intuitive result because more topological differences can not lead to larger Chernoff information between Gaussian trees.

\begin{table*}[!t]
  \centering
  \begin{tabular}{|c|c|c|c|c|c|}
    \hline
    Cases&$\lambda^*_{T_1||T_3}$&$\lambda$&
    $CI_{T_1||T_3}$&$CI_{T_1||T_2}$&$CI_{T_2||T_3}$\\
    \hline
     1&0.5191&$\begin{matrix}19.5746,
   0.0433,
   1.5439,
   0.7642,
   1,
   1,
   1\end{matrix}$&0.8983&0.9142  &  0.0251\\
     \hline
     2&0.5073&$\begin{matrix}9.2341,
    0.1019,
    1.2982,
    0.8185,
    1,
    1,
    1\end{matrix}$&0.5402&0.5418  &  0.0113 \\ \hline
    3&0.5254&$\begin{matrix}9.4328,
    1.653,
    0.0844,
    0.7603,
    1,
    1,
    1\end{matrix}$&0.5982&0.6103  &  0.0392 \\
    \hline
     4&0.5082
&$\begin{matrix}5.0195,
    0.1863,
    1.2201,
    0.8766,
    1,
    1,
    1\end{matrix}$&0.3102&0.3132&0.0056   \\
    \hline
    \end{tabular}
  \caption{Numerical cases dissatisfying Proposition \ref{<} in the chain of Fig. \ref{12}}\label{form2}
\end{table*}

\begin{figure}
  \centering
  \includegraphics[width=8cm]{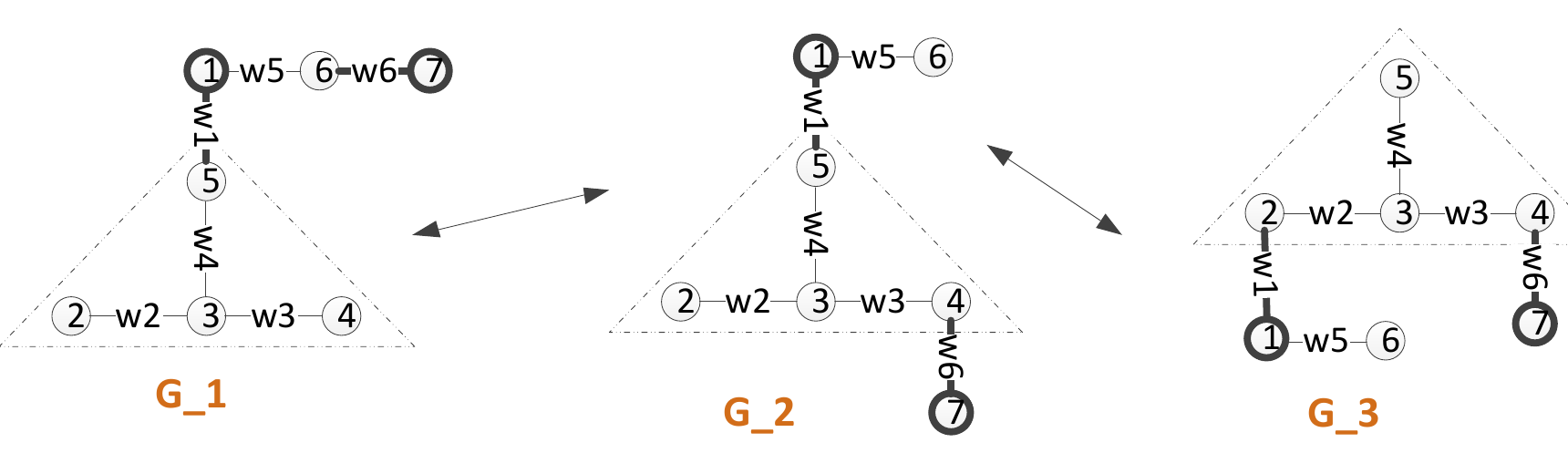}\\
  \caption{Example for dependent grafting operations}\label{12}
\end{figure}

\section{Dimension reduction}\label{s5}

    The situations in the former section are all about full observation cases, where we can observe all Gaussian variables in the trees each time.
    In practice, we may have some constraints on observation costs, which prompts us to reduce the dimension of observation vectors  in order to meet such  constraints. We can only use linearly transformed low-dimensional samples to do the classification. The linear transformation matrix should make sure that the reduced data have the maximum information for classification.

In section \ref{3A}, we have shown Proposition \ref{prop0}.
{With this proposition, we can inversely and linearly transform two original Gaussian graphs into isolated node graphs of new linear space. That is to say, variables of two distributions in new space are independent from each other. In this way, we can decompose difference information into independent dimensions.} {After decomposition, we can choose the dimensions with most difference information as classification dimension reduction result. The choice of dimensions is shown as follows.}

In these new space, $x_i'$, the $i$-th variable of $\mathbf{x}'=\mathbf{P}\mathbf{x}$, follows $N(0,1)$ in hypothesis $2$ and $N(0,\lambda_i)$ in hypothesis $1$. If $\lambda_i$ is farther from $1$, this dimension can provide more information for classification than other dimensions.

Assume that $m$ of all the $N$ eigenvalues $\{\lambda_i\}$ are greater than $1$ and the other $N-m$ eigenvalues are no more than $1$.
If we want to reduce the observation dimension from $N$ to $N_O$, we choose the dimensions of $\mathbf{x}'$ corresponding to the first $k$ rank and last $N_O-k$ rank of $\{\lambda_i\}$, where $\max\{N_O+m-N,0\}\leq k\leq \min\{m, N_O\}$. The $N_O\times N$ linear transformation matrix $\mathbf{A}_k$ is the corresponding $N_O$ rows of $\mathbf{P}$ corresponding to the chosen eigenvalues. 

Matrices $\mathbf{A}_k$ are candidate matrices of optimal classification linear dimension reduction matrix  $\mathbf{A}^*$, as shown below.

\begin{prop}\label{DR}
    Matrix $\mathbf{A}^*$ is the optimal $N_O\times N$ linear transformation matrix to maximize the Chernoff information in transformed space, namely
    $\mathbf{A}^*=\arg \max_{\mathbf{A}_{N_O\times N}} CI(\hat{\mathbf{\Sigma}}_1||\hat{\mathbf{\Sigma}}_2)$ where $\hat{\mathbf{\Sigma}}_i=\mathbf{A}\mathbf{\Sigma}_i{\mathbf{A}}^T$ for $i=1,2$. 
\begin{align}\mathbf{A}^*\in\{\mathbf{A}_k|\max\{N_O+m-N,0\}\leq k\leq \min\{m, N_O\}\}\end{align}
    \end{prop}

Proof of proposition \ref{DR} can be seen in Appendix \ref{A7} and this proposition ensure the optimality of our method.

The observation is $\mathbf{y}=\mathbf{A}^*\mathbf{x}$ and the covariance matrices of $\mathbf{y}$ in two hypotheses are
    \begin{align}
\mathbf{\Sigma}_2''=\mathbf{A}^*\mathbf{\Sigma}_2{\mathbf{A}^*}^T=\mathbf{I}_{N_O}\\
\mathbf{\Sigma}_1''=\mathbf{A}^*\mathbf{\Sigma}_1{\mathbf{A}^*}^T=Diag(\{\mu_i\})
\end{align}
where $\mathbf{\Sigma}_1''$ and $\mathbf{\Sigma}_2''$ are $N_O\times N_O$ diagonal matrices and $\{\mu_1,\mu_2,\dots,\mu_{N_O}\}$ (including multiple eigenvalues) are $N_O$ chosen eigenvalues.

The main thinking of our method is similar with PCA. We first decompose the information into independent dimensions and then choose the dimensions with largest weights. But the methods by which we decompose information and choose
dimensions are quite different.

\section{Conclusion}\label{s6}

In this paper, we show the relationship between topology, statistical
distribution and algebra.  Chernoff information between two Gaussian graphs can be determined by the generalized eigenvalues of their covariance matrices. Unit generalized eigenvalues are very special and do not affect the Chernoff information. Adding and division operations on Gaussian trees only add a newly unit generalized eigenvalues and do not change other generalized eigenvalues. Thus these operations keep the Chernoff information. We also extend our former result about grafting operation to Gaussian trees connected by more than one independent grafting operation and provide a partial ordering among these trees. In addition, we provide an optimal classification linear dimension reduction method with the metric of Chernoff information.

\bibliography{Bibliography}

\begin{thebibliography}{10}
\providecommand{\url}[1]{#1}
\csname url@samestyle\endcsname
\providecommand{\newblock}{\relax}
\providecommand{\bibinfo}[2]{#2}
\providecommand{\BIBentrySTDinterwordspacing}{\spaceskip=0pt\relax}
\providecommand{\BIBentryALTinterwordstretchfactor}{4}
\providecommand{\BIBentryALTinterwordspacing}{\spaceskip=\fontdimen2\font plus
\BIBentryALTinterwordstretchfactor\fontdimen3\font minus
  \fontdimen4\font\relax}
\providecommand{\BIBforeignlanguage}[2]{{%
\expandafter\ifx\csname l@#1\endcsname\relax
\typeout{** WARNING: IEEEtran.bst: No hyphenation pattern has been}%
\typeout{** loaded for the language `#1'. Using the pattern for}%
\typeout{** the default language instead.}%
\else
\language=\csname l@#1\endcsname
\fi
#2}}
\providecommand{\BIBdecl}{\relax}
\BIBdecl

\bibitem{vega2007complex}
F.~Vega-Redondo, \emph{Complex social networks}.\hskip 1em plus 0.5em minus
  0.4em\relax Cambridge University Press, 2007, no.~44.

\bibitem{dobra2010modeling}
A.~Dobra, T.~S. Eicher, and A.~Lenkoski, ``Modeling uncertainty in
  macroeconomic growth determinants using {Gaussian} graphical models,''
  \emph{Statistical Methodology}, vol.~7, no.~3, pp. 292--306, 2010.

\bibitem{ahmed2008time}
\BIBentryALTinterwordspacing
A.~Ahmed, L.~Song, and E.~P. Xing, ``Time-varying networks: Recovering
  temporally rewiring genetic networks during the life cycle of drosophila
  melanogaster,'' \emph{arXiv}, 2008. [Online]. Available:
  \url{https://arxiv.org/abs/0901.0138}
\BIBentrySTDinterwordspacing

\bibitem{cover2012elements}
T.~M. Cover and J.~A. Thomas, \emph{Elements of information theory}.\hskip 1em
  plus 0.5em minus 0.4em\relax John Wiley \& Sons, 2012.

\bibitem{chernoff1952measure}
H.~Chernoff, ``A measure of asymptotic efficiency for tests of a hypothesis
  based on the sum of observations,'' \emph{The Annals of Mathematical
  Statistics}, pp. 493--507, 1952.

\bibitem{westover2008asymptotic}
M.~B. Westover, ``Asymptotic geometry of multiple hypothesis testing,''
  \emph{IEEE Transactions on Information Theory}, vol.~54, no.~7, pp.
  3327--3329, 2008.

\bibitem{5540052}
G.~Doretto and Y.~Yao, ``Region moments: Fast invariant descriptors for
  detecting small image structures,'' in \emph{2010 IEEE Computer Society
  Conference on Computer Vision and Pattern Recognition}, 2010, pp. 3019--3026.

\bibitem{6247965}
R.~Wang, H.~Guo, L.~S. Davis, and Q.~Dai, ``Covariance discriminative learning:
  A natural and efficient approach to image set classification,'' in \emph{2012
  IEEE Conference on Computer Vision and Pattern Recognition}, 2012, pp.
  2496--2503.

\bibitem{binglin}
B.~Li, S.~Wei, Y.~Wang, and J.~Yuan, ``Chernoff information of bottleneck
  {Gaussian} trees,'' in \emph{2016 IEEE International Symposium on Information
  Theory (ISIT)}.\hskip 1em plus 0.5em minus 0.4em\relax IEEE, 2016, pp.
  970--974.

\bibitem{NOWAKOWSKA201674}
E.~Nowakowska, J.~Koronacki, and S.~Lipovetsky, ``Dimensionality reduction for
  data of unknown cluster structure,'' \emph{Information Sciences}, vol. 330,
  pp. 74 -- 87, 2016.

\bibitem{GUAN2013147}
H.~Guan, J.~Zhou, B.~Xiao, M.~Guo, and T.~Yang, ``Fast dimension reduction for
  document classification based on imprecise spectrum analysis,''
  \emph{Information Sciences}, vol. 222, pp. 147 -- 162, 2013.

\bibitem{6472238}
Y.~Bengio, A.~Courville, and P.~Vincent, ``Representation learning: A review
  and new perspectives,'' \emph{IEEE Transactions on Pattern Analysis and
  Machine Intelligence}, vol.~35, no.~8, pp. 1798--1828, 2013.

\end{thebibliography}
\bibliographystyle{IEEEtran}

\newpage
\appendices
\section{Proof of Proposition~\ref{prop0}}\label{A1}

Because $\mathbf{\Sigma}_1$ and $\mathbf{\Sigma}_2^{-1}$ are real symmetric positive definite matrices, $\mathbf{\Sigma}_1=\mathbf{L}_1\mathbf{L}_1^T$ and $\mathbf{\Sigma}_2^{-1}=\mathbf{L}_2\mathbf{L}_2^T$ due to Cholesky decomposition where $\mathbf{L}_1$ and $\mathbf{L}_2$ are real non-singular triangular matrices.

The characteristic polynomial of $\mathbf{\Sigma}_1\mathbf{\Sigma}_2^{-1}$ is
\begin{align}
f(\mathbf{\Sigma}_1\mathbf{\Sigma}_2^{-1})&=|\lambda \mathbf{I}-\mathbf{\Sigma}_1\mathbf{\Sigma}_2^{-1}|\nonumber\\&=
|\lambda \mathbf{I}-\mathbf{L}_1\mathbf{L}_1^T\mathbf{L}_2\mathbf{L}_2^T|\nonumber\\&=
|\mathbf{L}_1(\lambda \mathbf{I}-\mathbf{L}_1^T\mathbf{L}_2\mathbf{L}_2^T\mathbf{L}_1)\mathbf{L}_1^{-1}|\nonumber\\&=
|\lambda \mathbf{I}-\mathbf{L}_1^T\mathbf{L}_2\mathbf{L}_2^T\mathbf{L}_1|\nonumber\\&=
|\lambda \mathbf{I}-(\mathbf{L}_1^T\mathbf{L}_2){(\mathbf{L}_1^T\mathbf{L}_2)}^T|
\end{align}
We can conclude $\mathbf{\Sigma}_1\mathbf{\Sigma}_2^{-1}$ has the same eigenvalues as $(\mathbf{L}_1^T\mathbf{L}_2){(\mathbf{L}_1^T\mathbf{L}_2)}^T$, which is a real symmetric positive definite matrix. The eigenvalues of $\mathbf{\Sigma}_1\mathbf{\Sigma}_2^{-1}$ are all positive and $\mathbf{\Sigma}_1\mathbf{\Sigma}_2^{-1}$ is a positive definite matrix.

The eigenvalue decomposition of $\mathbf{\Sigma}_1\mathbf{\Sigma}_2^{-1}$ is $\mathbf{Q}\mathbf{\Lambda} \mathbf{Q}^{-1}$, where $\mathbf{Q}$ is an $N\times N$ matrix and $\mathbf{\Lambda}=Diag(\{\lambda_i\})$ is a diagonal matrix of eigenvalues, in which we put multiple eigenvalues adjacent. $\{\lambda_i\}$ are the eigenvalues of $\mathbf{\Sigma}_1\mathbf{\Sigma}_2^{-1}$, namely the generalized eigenvalues of $\mathbf{\Sigma}_1$ and $\mathbf{\Sigma}_2$.

New covariance matrices $\mathbf{\Sigma}_1^{(1)}=\mathbf{Q}^{-1}\mathbf{\Sigma}_1{\mathbf{Q}^{-1}}^T=[a_{ij}]$ and $\mathbf{\Sigma}_2^{(1)}=\mathbf{Q}^{-1}\mathbf{\Sigma}_2{\mathbf{Q}^{-1}}^T=[b_{ij}]$ satisfy
$\mathbf{\Sigma}_1^{(1)}=\mathbf{\Lambda} \mathbf{\Sigma}_2^{(1)}$.
\begin{align}
a_{ij}=\lambda_ib_{ij}\\
a_{ij}=a_{ji}=\lambda_jb_{ji}=\lambda_jb_{ij}
\end{align}
For $\forall i\neq j$, $a_{ij}=b_{ij}=0$ or $\lambda_i=\lambda_j$.

If the eigenvalues of $\mathbf{\Sigma}_1\mathbf{\Sigma}_2^{-1}$ have $k$ different multiple eigenvalues $\lambda^{(1)},\lambda^{(2)},\dots,\lambda^{(k)}$ with multiplicity $n_1,n_2,\dots,n_k$, so
\begin{align}
\mathbf{\Sigma}_2^{(1)}=\begin{bmatrix}
\mathbf{J}_1\\
&\mathbf{J}_2\\
&&\mathbf{J}_3\\
&&&\ddots\\
&&&&\mathbf{J}_k
\end{bmatrix}\\
\mathbf{\Sigma}_1^{(1)}=\begin{bmatrix}
\lambda^{(1)}\mathbf{J}_1\\
&\lambda^{(2)}\mathbf{J}_2\\
&&\lambda^{(3)}\mathbf{J}_3\\
&&&\ddots\\
&&&&\lambda^{(k)}\mathbf{J}_k
\end{bmatrix}
\end{align}
where $\mathbf{J}_i$ is $n_i\times n_i$ inverse matrix.

We can also get an inverse matrix $\mathbf{Q}^{(1)}={\mathbf{\Sigma}_2^{(1)}}^{-0.5}$ so that
\begin{align}
\mathbf{\Sigma}_2^{(2)}=\mathbf{Q}^{(1)}\mathbf{\Sigma}_2^{(1)}{\mathbf{Q}^{(1)}}^T=\mathbf{I}_N\\
\mathbf{\Sigma}_1^{(2)}=\mathbf{Q}^{(1)}\mathbf{\Sigma}_1^{(1)}{\mathbf{Q}^{(1)}}^T=\mathbf{\Lambda}
\end{align}

We can construct a linear transformation matrix
$\mathbf{P}=\mathbf{Q}^{(1)}\mathbf{Q}^{-1}={\left(\mathbf{Q}^{-1}\mathbf{\Sigma}_2{(\mathbf{Q}^{-1})}^T\right)}^{-\frac{1}{2}}\mathbf{Q}^{-1}$ and thus
\begin{align}
\mathbf{\Sigma}_2'=&\mathbf{P}\mathbf{\Sigma}_2\mathbf{P}^T=\mathbf{I}_N\\
\mathbf{\Sigma}_1'=&\mathbf{P}\mathbf{\Sigma}_1\mathbf{P}^T=\mathbf{\Lambda}
\end{align}

\section{Proof of Proposition~\ref{prop1}}\label{A2}

Parameter $\lambda^*$ satisfies $D(\mathbf{\Sigma}_{\lambda^*}'||\mathbf{\Sigma}_1')=D(\mathbf{\Sigma}_{\lambda^*}'||\mathbf{\Sigma}_2')$, and thus
\begin{align}
\frac{1}{2}&\sum_i\left(\ln\left(\lambda^*+(1-\lambda^*)\lambda_i\right)+\frac{1}{\lambda^*+(1-\lambda^*)\lambda_i}-1\right)\nonumber\\
=\frac{1}{2}&\sum_i\left(\ln\frac{\lambda^*+(1-\lambda^*)\lambda_i}{\lambda_i}+\frac{\lambda_i}{\lambda^*+(1-\lambda^*)\lambda_i}-1\right)
\end{align}
With $\prod_i \lambda_i=\beta$, we can conclude
\begin{align}
\ln\beta+\sum_i\frac{1}{\lambda^*+(1-\lambda^*)\lambda_i}=\sum_i\frac{\lambda_i}{\lambda^*+(1-\lambda^*)\lambda_i}
\end{align}

We also know
\begin{align}
\lambda^*\sum_i\frac{1}{\lambda^*+(1-\lambda^*)\lambda_i}+&(1-\lambda^*)\sum_i\frac{\lambda_i}{\lambda^*+(1-\lambda^*)\lambda_i}
\nonumber\\=\sum_i1=&N
\end{align}

Parameter $\lambda^*\in[0,1]$ is the unique result of
    \begin{align}
\sum_{i} \frac{1}{\lambda^*+(1-\lambda^*)\lambda_i}=N+(\lambda^*-1)\ln\beta\\
\sum_{i} \frac{\lambda_i}{\lambda^*+(1-\lambda^*)\lambda_i}=N+\lambda^*\ln\beta
    \end{align}

\begin{align}
    &CI(\mathbf{\Sigma}_1||\mathbf{\Sigma}_2)=D(\mathbf{\Sigma}_{\lambda^*}'||\mathbf{\Sigma}_1')
\nonumber\\=&\frac{1}{2}\sum_i\left(\ln\left(\lambda^*+(1-\lambda^*)\lambda_i\right)+\frac{1}{\lambda^*+(1-\lambda^*)\lambda_i}-1\right)
\nonumber\\=&\frac{1}{2} \sum_i \ln\left((1-\lambda^*) \sqrt{\lambda_i}+\frac{\lambda^*}{\sqrt{\lambda_i}}\right)+ \frac{1}{2}(\lambda^*- \frac{1}{2})\ln\beta
    \end{align}

\section{Proof of Proposition~\ref{prop3}}\label{A3}

Parameter $\lambda^*$ is unique in the range of $[0,1]$. If $\lambda^*$ satisfies equation (\ref{1}), then
    \begin{align}
   & \sum_{i} \frac{\lambda_i}{\lambda^*+(1-\lambda^*)\lambda_i}+\frac{1}{\lambda^*+(1-\lambda^*)}\nonumber\\=&N+1+\lambda^*\ln\beta
    \end{align} where $\{\lambda_i\}$ are the generalized eigenvalues of $(\mathbf{\Sigma}_1,\mathbf{\Sigma}_2)$.

    The generalized eigenvalues of $(\mathbf{\Sigma}_1',\mathbf{\Sigma}_2')$ are $\{\lambda_i\}\bigcup\{1\}$ and $\lambda^*$ also satisfies equation (\ref{1}). The optimal $\lambda^*$ is the same. 
    \begin{align}
    &CI(\mathbf{\Sigma}_1'||\mathbf{\Sigma}_2')\nonumber\\
=&
    \frac{1}{2} \sum_{i=1}^N \left\{\ln\{(1-\lambda^*) \sqrt{\lambda_i}+\frac{\lambda^*}{\sqrt{\lambda_i}}\}\right\}+ \frac{1}{2}(\lambda^*- \frac{1}{2})\ln\beta\nonumber\\
    +&\frac{1}{2} \left\{\ln\{(1-\lambda^*) \sqrt{1}+\frac{\lambda^*}{\sqrt{1}}\}\right\}
\nonumber\\
    =&\frac{1}{2} \sum_{i=1}^N \left\{\ln\{(1-\lambda^*) \sqrt{\lambda_i}+\frac{\lambda^*}{\sqrt{\lambda_i}}\}\right\}+ \frac{1}{2}(\lambda^*- \frac{1}{2})\ln\beta\nonumber\\
    =&CI(\mathbf{\Sigma}_1||\mathbf{\Sigma}_2)
    \end{align}

\section{Proof of Proposition~\ref{prop2}}\label{A4}

    We use the equation of block matrix to prove this proposition. We deal with adding operation and division operation in different subsections.

\subsection{Proof on adding operation case}

    Adding operation is shown in Fig. {\ref{adding}}. Assume $G_1$ and $G_2$ have node $1,\dots,N$, and their covariance matrices are

    $G_1$: $\mathbf{\Sigma}_1=\begin{bmatrix} \mathbf{D}&\boldsymbol{\alpha}\\ \boldsymbol{\alpha}^T&1 \end{bmatrix}$

    $G_2$: $\mathbf{\Sigma}_2$, $\mathbf{\Sigma}_2^{-1}=\begin{bmatrix} \mathbf{A}_2&\mathbf{B}_2\\\mathbf{B}_2^T&a_2 \end{bmatrix}$.

     Without loss of generality, the new edge is $e_{N,N+1}$ with parameter $w$. New covariance matrices after adding operation are
     \begin{align}
     \mathbf{\Sigma}_1'=\begin{bmatrix} \mathbf{D}&\boldsymbol\alpha&w\boldsymbol\alpha\\ \boldsymbol\alpha^T&1&w\\w\boldsymbol\alpha^T&w&1 \end{bmatrix}\\
     \mathbf{\Sigma}_2'^{-1}=
     \begin{bmatrix} \mathbf{A}_2&\mathbf{B}_2&0\\\mathbf{B}_2^T&a_2+\frac{w^2}{1-w^2}&\frac{-w}{1-w^2}\\
     0&-\frac{w}{1-w^2}&\frac{1}{1-w^2} \end{bmatrix}
     \end{align}

     The generalized eigenvalues of $\mathbf{\Sigma}_1$ and $\mathbf{\Sigma}_2$ are the roots of $|\lambda \mathbf{I} -\mathbf{\Sigma}_1\mathbf{\Sigma}_2^{-1}|=0$.

     \begin{align}
     \mathbf{\Sigma}_1\mathbf{\Sigma}_2^{-1}=&\begin{bmatrix} \mathbf{D}&\boldsymbol\alpha\\ \boldsymbol\alpha^T&1 \end{bmatrix}\begin{bmatrix} \mathbf{A}_2&\mathbf{B}_2\\\mathbf{B}_2^T&a_2 \end{bmatrix}\nonumber\\
     =&
     \begin{bmatrix} \mathbf{D}\mathbf{A}_2+\boldsymbol\alpha \mathbf{B}_2^T&\mathbf{D}\mathbf{B}_2+a_2\boldsymbol\alpha\\ \boldsymbol\alpha^T\mathbf{A}_2+\mathbf{B}_2^T& \boldsymbol\alpha^T\mathbf{B}_2+a_2 \end{bmatrix}\\
     \mathbf{\Sigma}_1'\mathbf{\Sigma}_2'^{-1}=&
     \begin{bmatrix}
     \mathbf{D}\mathbf{A}_2+\boldsymbol\alpha \mathbf{B}_2^T&\mathbf{D}\mathbf{B}_2+a_2\boldsymbol\alpha&0\\
     \boldsymbol\alpha^T\mathbf{A}_2+\mathbf{B}_2^T& \boldsymbol\alpha^T\mathbf{B}_2+a_2 &0\\
     \ast&\ast&1
     \end{bmatrix}\\
     |\lambda \mathbf{I} -\mathbf{\Sigma}_1'\mathbf{\Sigma}_2'^{-1}|=&(\lambda-1)|\lambda \mathbf{I } -\mathbf{\Sigma}_1\mathbf{\Sigma}_2^{-1}|
     \end{align}
where $\ast$ are matrices that do not affect our result.
     New trees have an extra unit generalized eigenvalue without changing other eigenvalues.

\subsection{Proof on division operation case}

    Division operation is shown in Fig. {\ref{division}}. Assume that $G_1$ and $G_2$ have node $1,\dots,N$,  and the connecting points are $p=1,q=N$ without loss of generality.

    If we set edge $e_{1N}$ in $G_2$ to be $0$, we can treat it as a new $N$-node tree, whose covariance matrix is $\mathbf{A}^{-1}$.
    We divide $\mathbf{A}$ as
    \begin{align}
    \mathbf{A}=\begin{bmatrix} x&\mathbf{X}^T&0\\\mathbf{X}&\mathbf{Z}&\mathbf{Y}\\0&\mathbf{Y}^T&y \end{bmatrix}
    \end{align}
    where $\mathbf{X},\mathbf{Y}$ are $(N-2)\times1$ matrices and $\mathbf{Z}$ is $(N-2)\times(N-2)$ matrix.

    In $G_1$ and $G_1'$, we define the column node $1$ to nodes $2$ to $N-1$ to be $\boldsymbol\alpha_1$(setting the respective rows to the nodes in $G_{T_{12}}$ $0$) and the column node $N$ to nodes $2$ to $N-1$ to be $\boldsymbol\alpha_2$ (setting the respective rows to the nodes in $G_{T_{11}}$ $0$). Then in $G_1$ and $G_1'$, the column from node $1$ to nodes $2$ to $N-1$ is $\boldsymbol\alpha_1+w_1w_2\boldsymbol\alpha_2$ because node $1$ can reach the nodes in $G_{T_{11}}$ directly, but reach the nodes in $G_{T_{12}}$ through node $N$ with path weighted by $w_{1N}=w_1w_2$. The column from node $N$ to node $2$ to $N-1$ is $w_1w_2\boldsymbol\alpha_1+\boldsymbol\alpha_2$ because node $N$ can reach the nodes in $G_{T_{12}}$ directly, but reach the nodes in $G_{T_{11}}$ through node $1$ with path weighted by $w_{1N}=w_1w_2$. In $G_1'$, the column from node $N+1$ to node $2$ to $N-1$ is $w_1\boldsymbol\alpha_1+w_2\boldsymbol\alpha_2$ because node $N+1$ can reach the nodes in $G_{T_{11}}$ through node $1$ with path $e_{1,N+1}$ weighted by $w_{1}$, and reach the nodes in $G_{T_{12}}$ through node $N$ with path $e_{N,N+1}$ weighted by $w_{2}$.

    The covariance matrices of these trees are as shown below:
\scriptsize
\begin{equation}
G_1:~\mathbf{\Sigma}_{11}=\begin{bmatrix}
1&\boldsymbol\alpha_1^T+w_1w_2\boldsymbol\alpha_2^T&w_1w_2\\
    \boldsymbol\alpha_1+w_1w_2\boldsymbol\alpha_2 &\mathbf{P}&w_1w_2\boldsymbol\alpha_1+\boldsymbol\alpha_2 \\
    w_1w_2&w_1w_2\boldsymbol\alpha_1^T+\boldsymbol\alpha_2^T &1
\end{bmatrix}
\end{equation}
\begin{equation}
G_2: \mathbf{\Sigma}_{12}, \mathbf{\Sigma}_{12}^{-1}=\begin{bmatrix}
 x+\frac{w_1^2w_2^2}{1-w_1^2w_2^2}&\mathbf{X}^T&-\frac{w_1w_2}{1-w_1^2w_2^2}\\\mathbf{X}&\mathbf{Z}&\mathbf{Y}\\-\frac{w_1w_2}{1-w_1^2w_2^2}&\mathbf{Y}^T&y+\frac{w_1^2w_2^2}{1-w_1^2w_2^2}
\end{bmatrix}
\end{equation}
\begin{align}
&G_1':  
\mathbf{\Sigma}_{21}\nonumber\\
=&\begin{bmatrix}
1&\boldsymbol\alpha_1^T+w_1w_2\boldsymbol\alpha_2^T&w_1w_2&w_1\\
   \boldsymbol \alpha_1+w_1w_2\boldsymbol\alpha_2 &\mathbf{P}&w_1w_2\boldsymbol\alpha_1+\boldsymbol\alpha_2 &w_1\boldsymbol\alpha_1+w_2\boldsymbol\alpha_2\\
    w_1w_2&w_1w_2\boldsymbol\alpha_1^T+\boldsymbol\alpha_2^T &1&w_2\\
    w_1&w_1\boldsymbol\alpha_1^T+w_2\boldsymbol\alpha_2^T&w_2&1
\end{bmatrix}
\end{align}
\begin{equation}
G_2': \mathbf{\Sigma}_{22}, \mathbf{\Sigma}_{22}^{-1}=\begin{bmatrix}
x+\frac{w_1^2}{1-w_1^2}&\mathbf{X}^T&\mathbf{0}&-\frac{w_1}{1-w_1^2}\\
    \mathbf{X}&\mathbf{Z}&\mathbf{Y}&\boldsymbol0\\
    \boldsymbol0&\mathbf{Y}^T&y+\frac{w_2^2}{1-w_2^2}&-\frac{w_2}{1-w_2^2}\\
    -\frac{w_1}{1-w_1^2}&\boldsymbol0&-\frac{w_2}{1-w_2^2}&\frac{1}{1-w_1^2}+\frac{w_2^2}{1-w_2^2}
\end{bmatrix}
\end{equation}
\normalsize

     The generalized eigenvalues of $\mathbf{\Sigma}_1$ and $\mathbf{\Sigma}_2$ are the roots of $|\lambda \mathbf{I} -\mathbf{\Sigma}_1\mathbf{\Sigma}_2^{-1}|=0$.
     \begin{align}
     \mathbf{\Sigma}_{11}\mathbf{\Sigma}_{12}^{-1}=&\mathbf{\Sigma}_{11}\mathbf{A}+
     \begin{bmatrix} 0&0&-w_1w_2\\
     -w_1w_2\boldsymbol\alpha_2&0&-w_1w_2\boldsymbol\alpha_1\\
     -w_1w_2&0&0 \end{bmatrix}\\
     \mathbf{\Sigma}_{21}\mathbf{\Sigma}_{22}^{-1}=&
     \begin{bmatrix}
     \mathbf{\Sigma}_{11}\mathbf{\Sigma}_{12}^{-1}&\boldsymbol{0}\\
     \ast&1
     \end{bmatrix}\\
     |\lambda \mathbf{I} -\mathbf{\Sigma}_{21}\mathbf{\Sigma}_{22}^{-1}|=&(\lambda-1)|\lambda \mathbf{I} -\mathbf{\Sigma}_{11}\mathbf{\Sigma}_{12}^{-1}|
     \end{align}
     New trees have an extra unit generalized eigenvalue  without changing other eigenvalues.

\section{Proof of Proposition~\ref{prop7}}\label{A5}

   Parameter $\lambda^*$ satisfies $D(\mathbf{\Sigma}_{\lambda^*}||\mathbf{\Sigma}_2)=D(\mathbf{\Sigma}_{\lambda^*}||\mathbf{\Sigma}_1)$. When $|\mathbf{\Sigma}_1|=|\mathbf{\Sigma}_2|$, it will become the unique root in $[0,1]$ of
    $tr\left(\mathbf{\Sigma}_{\lambda^*}( \mathbf{\Sigma}_1^{-1}- \mathbf{\Sigma}_2^{-1})\right)=0$. We only need to prove $tr\left(\mathbf{\Sigma}_{0.5}( \mathbf{\Sigma}_1^{-1}- \mathbf{\Sigma}_2^{-1})\right)=0$ before we get $\lambda^*=1/2$.

    $tr\left(\mathbf{A}\mathbf{B}\right)=\sum_{i,j} a_{ij}b_{ij}$ when $\mathbf{A},\mathbf{B}$ are both symmetric matrices. Luckily, $\mathbf{A}=\mathbf{\Sigma}_{0.5}$ and $\mathbf{B}=\mathbf{\Sigma}_1^{-1}- \mathbf{\Sigma}_2^{-1}$ are both symmetric matrices. In addition, $\mathbf{B}=\mathbf{\Sigma}_1^{-1}- \mathbf{\Sigma}_2^{-1}$ only have $2n$ non-zero diagonal elements and $2n$ pairs of non-zero non-diagonal elements, where $n$ is the number of grafting operations. For example, $b_{pp}=\frac{w^2}{1-w^2},b_{qq}=-\frac{w^2}{1-w^2},
    b_{iq}=\frac{w}{1-w^2},b_{ip}=-\frac{w}{1-w^2}$ if we cut node $i$ from node $p$ with edge $e_{ip}$ weighting $w$ and paste it to node $q$,
and
there is no other grafting operation involving these three nodes.
    Equation $tr\left(\mathbf{\Sigma}_{0.5}( \mathbf{\Sigma}_1^{-1}- \mathbf{\Sigma}_2^{-1})\right)$ is a sum of $4n$ terms, each $4$ terms correspond to one grafting operation. We can divide the polynomial into $n$ parts corresponding to each grafting operation, namely, $tr\left(\mathbf{\Sigma}_{0.5}( \mathbf{\Sigma}_1^{-1}- \mathbf{\Sigma}_2^{-1})\right)=\sum_k P_k$. Then we only need to prove each $P_t$ equals to $0$.

    As Fig. \ref{z} shows, there are three different types of grafting operations here.
    After simplification by using the inverse operations of adding and division operations, the three types of subparts can be translated as shown in Fig. \ref{10}. In this figure, we label these relevant nodes as $1,2,3\dots$ for simplification because it does not change the result if we exchange labels of nodes.

    \begin{figure}
  \centering
  \includegraphics[width=8cm]{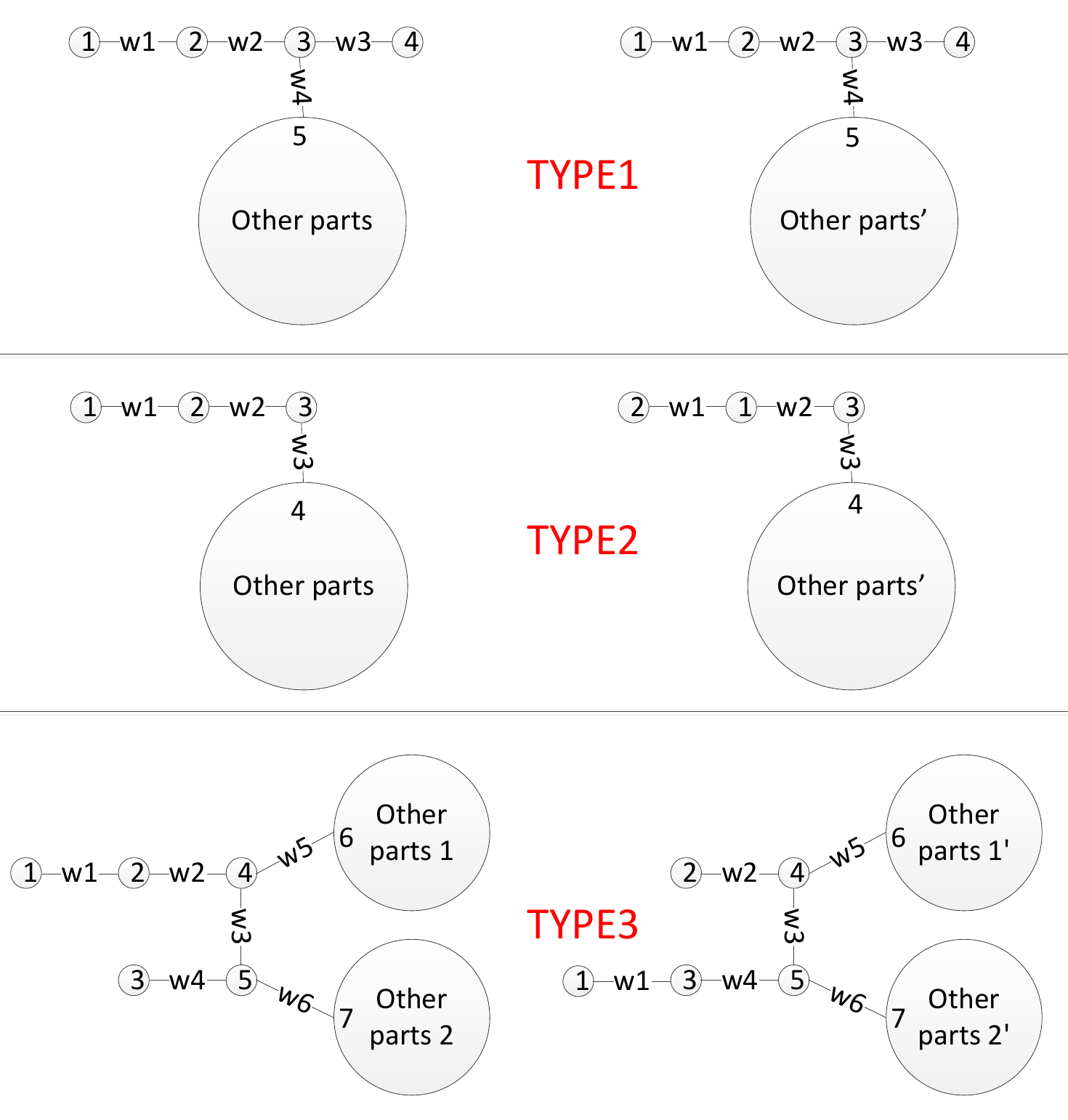}\\
  \caption{Simplified separate grafting operation types}\label{10}
  \end{figure}

    Before we deal with $P_k$ in each case, we show $\mathbf{\Sigma}_{\lambda}^{-1}$ is positive definite matrix. $\mathbf{\Sigma}_1$ and $\mathbf{\Sigma}_2$ are positive definite matrices, so $\mathbf{\Sigma}_1^{-1}$ and $\mathbf{\Sigma}_2^{-1}$ are positive definite matrices, namely  $\mathbf{z}^T\mathbf{\Sigma}_1^{-1} \mathbf{z}>0$ and $\mathbf{z}^T\mathbf{\Sigma}_2^{-1} \mathbf{z}>0$ for any non-zero vector $\mathbf{z}$. Then $\mathbf{z}^T\mathbf{\Sigma}_{\lambda}^{-1}\mathbf{z}=\lambda \mathbf{z}^T\mathbf{\Sigma}_1^{-1}\mathbf{z}+\mathbf{z}^T(1-\lambda)\mathbf{\Sigma}_2^{-1}\mathbf{z}>0$ for any non-zero vector $\mathbf{z}$. Matrix $\mathbf{\Sigma}_{\lambda}^{-1}$ is positive definite matrix and its order principal minors are invertible matrices. This result is useful for our later proofs.

    We define $f_i=\frac{-w_i}{1-w_i^2}$ and $g_i=\frac{w_i^2}{1-w_i^2}$, so $(1+g_i)g_i=f^2_i$. Then
    we  prove $P_k=0$ in different type of Fig. \ref{10} later.

\subsection{TYPE1}

In this type, we focus on node $1$ to $4$, which is involved in this grafting operation. 

$\mathbf{\Sigma}_1^{-1}=$
\[
  \begin{pmat}[{...|}]
  1+g_1&f_1&&&\cr
    f_1&1+g_1+g_2&f_2&&\cr
    &f_2&1+g_2+g_3+g_4&f_3&\mathbf{V} \cr
    &&f_3&1+g_3&\cr\-
    &&\mathbf{V}^T&&\mathbf{K^{(1)}}\cr
  \end{pmat}
\]
$\mathbf{\Sigma}_2^{-1}=$
\[
  \begin{pmat}[{...|}]
  1+g_1&&&f_1&\cr
    &1+g_2&f_2&&\cr
    &f_2&1+g_2+g_3+g_4&f_3&\mathbf{V} \cr
    f_1&&f_3&1+g_1+g_3&\cr\-
    &&\mathbf{V}^T&&\mathbf{K^{(2)}}\cr
  \end{pmat}
\]
where $\mathbf{V}$ is a $4\times (N-4)$ matrix with all zero elements except $v_{3,1}=f_4$ and $\mathbf{K^{(1)}},\mathbf{K^{(2)}}$ are covariance matrices of node $5$ to $N$ in $G_1$ and $G_2$ respectively.

$\mathbf{\Sigma}_1^{-1}-\mathbf{\Sigma}_2^{-1}=$
\[
  \begin{pmat}[{...|}]
  0&f_1&&-f_1&\cr
    f_1&g_1&&&\cr
    &&0&& \cr
    -f_1&&&-g_1&&\cr\-
    &&&&\mathbf{K^{(1)}}-\mathbf{K^{(2)}}\cr
  \end{pmat}
\]
$\mathbf{\Sigma}_{0.5}^{-1}=\frac{1}{2}\mathbf{\Sigma}_1^{-1}+\frac{1}{2}\mathbf{\Sigma}_2^{-1}=$\tiny
\[
  \begin{pmat}[{...|}]
  1+\frac{1}{2}g_1&\frac{1}{2}f_1&&\frac{1}{2}f_1&\cr
    \frac{1}{2}f_1&1+g_2+\frac{1}{2}g_1&f_2&&\cr
    &f_2&1+g_2+g_3+g_4&f_3&\mathbf{V}\cr
    \frac{1}{2}f_1&&f_3&1+\frac{1}{2}g_1+g_3&\cr\-
    &\mathbf{V}^T&&&\mathbf{K^{(3)}}\cr
  \end{pmat}
\]\normalsize
where $\mathbf{K^{(3)}}=\frac{1}{2}\mathbf{K^{(1)}}+\frac{1}{2}\mathbf{K^{(2)}}$
     is an invertible matrix because $\mathbf{\Sigma}_{\lambda}^{-1}$ is positive definite matrix. We define
    \begin{align}
    \mathbf{V}{\mathbf{K^{(3)}}}^{-1}\mathbf{V}^T=\begin{bmatrix}
    0&&&\\&0&&\\&&\frac{X}{|\mathbf{K^{(3)}}|}&\\&&&0
    \end{bmatrix}\end{align}

    We can conclude $P_k$ related to this grafting operation is $g_1m_{22}-g_1m_{44}+2f_1m_{12}-2f_1m_{14}$ where $m_{ij}$ is the $ij$-th element of $\mathbf{\Sigma}_{0.5}$.

    $M_{ij}$ is the $(i,j)$ minor of $\mathbf{\Sigma}_{0.5}^{-1}$, so
    \begin{align}
    M_{22}=&|\mathbf{K^{(3)}}|(1+g_1)\{(1+g_3)(1+g_2+g_4+\frac{1}{4}g_1)\nonumber\\+&\frac{1}{4}g_1(g_2+g_4)\}-(1+g_1)(1+g_3+\frac{1}{4}g_1)X\nonumber\\
    M_{44}=&|\mathbf{K^{(3)}}|(1+g_1)\{(1+g_2)(1+g_3+g_4+\frac{1}{4}g_1)\nonumber\\+&\frac{1}{4}g_1(g_3+g_4)\}-(1+g_1)(1+g_2+\frac{1}{4}g_1)X\nonumber\\
    M_{12}=&|\mathbf{K^{(3)}}|\{\frac{1}{2}f_1\{(1+g_3)(1+g_2+g_4+\frac{1}{2}g_1)\nonumber\\+&
    \frac{1}{2}g_1(g_2+g_4)\}+\frac{1}{2}f_1f_2f_3\}
    -\frac{1}{2}f_1(1+g_3+\frac{1}{2}g_1)X\nonumber\\
    M_{14}=&|\mathbf{K^{(3)}}|\{\frac{1}{2}f_1\{(1+g_2)(1+g_3+g_4+\frac{1}{2}g_1)\nonumber\\+&
    \frac{1}{2}g_1(g_3+g_4)\}+\frac{1}{2}f_1f_2f_3\}
    -\frac{1}{2}f_1(1+g_2+\frac{1}{2}g_1)X\nonumber
    \end{align}
    and
    \begin{align}
    &2f_1(-M_{12}+M_{14})+g_1(M_{22}-M_{44})=0\\
    P_k=&g_1m_{22}-g_1m_{44}+2f_1m_{12}-2f_1m_{14}\nonumber\\=&
    \left(2f_1(-M_{12}+M_{14})+g_1(M_{22}-M_{44})\right)|\mathbf{\Sigma}_{0.5}|=0
    \end{align}

\subsection{TYPE2}

In this type, we focus on node $1$ to $3$, which is involved in this grafting operation.  

$\mathbf{\Sigma}_1^{-1}=$
\[
  \begin{pmat}[{..|}]
  1+g_1&f_1&&\cr
    f_1&1+g_1+g_2&f_2&\mathbf{V} \cr
    &f_2&1+g_2+g_3&\cr\-
    &\mathbf{V}^T&&\mathbf{K^{(1)}}\cr
  \end{pmat}
\]
$\mathbf{\Sigma}_2^{-1}=$
\[
  \begin{pmat}[{..|}]
  1+g_1+g_2&f_1&f_2&\cr
    f_1&1+g_1&&\mathbf{V}\cr
    f_2&&1+g_2+g_3&\cr\-
    &\mathbf{V}^T&&\mathbf{K^{(2)}}\cr
  \end{pmat}
\]
where $\mathbf{V}$ is a $3\times (N-3)$ matrix with all zero elements except $v_{3,1}=f_3$ and $\mathbf{K^{(1)}},\mathbf{K^{(2)}}$ are covariance matrices of node $4$ to $N$ in $G_1$ and $G_2$ respectively.
Then we use the same process and get
\begin{align}
    P_k=&g_2m_{22}-g_2m_{11}+2f_2m_{23}-2f_2m_{13}=0
\end{align}
    where $m_{ij}$ is the $ij$-th element of $\mathbf{\Sigma}_{0.5}$.

\subsection{TYPE3}

In this type, we focus on node $1$ to $5$, which is involved in this grafting operation.  

$\mathbf{\Sigma}_1^{-1}=$\tiny
\[
  \begin{pmat}[{....|}]
  1+g_1&f_1&&&&\cr
    f_1&1+g_1+g_2&&f_2&&\cr
    &&1+g_4&&f_4&\mathbf{V} \cr
    &f_2&&1+g_2+g_3+g_5&f_3&\cr
    &&f_4&f_3&1+g_3+g_4+g_6&\cr\-
    &&\mathbf{V}^T&&&\mathbf{K^{(1)}}\cr
  \end{pmat}
\]\normalsize
$\mathbf{\Sigma}_2^{-1}=$\tiny
\[
  \begin{pmat}[{....|}]
  1+g_1&&f_1&&&\cr
    &1+g_2&&f_2&&\cr
    f_1&&1+g_1+g_4&&f_4&\mathbf{V} \cr
    &f_2&&1+g_2+g_3+g_5&f_3&\cr
    &&f_4&f_3&1+g_3+g_4+g_6&\cr\-
    &&\mathbf{V}^T&&&\mathbf{K^{(2)}}\cr
  \end{pmat}
\]\normalsize
where $\mathbf{V}$ is a $5\times (N-5)$ matrix with all zero elements except $v_{4,1}=f_5,v_{5,2}=f_6$ and $\mathbf{K^{(1)}},\mathbf{K^{(2)}}$ are covariance matrices of node $6$ to $N$ in $G_1$ and $G_2$ respectively.
Then we use the same process and get
\begin{align}
    P_k=&g_1m_{22}-g_1m_{33}+2f_1m_{12}-2f_1m_{13}=0
\end{align}
    where $m_{ij}$ is the $ij$-th element of $\mathbf{\Sigma}_{0.5}$.

    $P_k=0$ in all types of grafting operations if these operations are separate.  $tr\left(\mathbf{\Sigma}_{0.5}( \mathbf{\Sigma}_1^{-1}- \mathbf{\Sigma}_2^{-1})\right)=\sum_k P_k=0$.

    Equation $\lambda^*=\frac{1}{2}$ holds if all the grafting operations in the chain are separate.

\section{Proof of Proposition~\ref{<}}\label{A6}

    We  prove this proposition by using adding and division operation  repeatedly.

    For $p\leq i\leq j\leq q$, there are $q-p$ grafting operations in the grafting chain $T_p \leftrightarrow \dots \leftrightarrow T_q$. We divide these grafting operations into two sets: grafting operations from $T_i$ to $T_j$ and other grafting operations, namely, set $1$ and set $2$.

    We can simplify tree pairs $(T_p,T_q)$ and  $(T_i,T_j)$ into $(\hat{T}_p,\hat{T}_q)$ and $(\tilde{T}_i,\tilde{T}_j)$  respectively by using the inverse operations of adding and division operation  repeatedly. $(\hat{T}_p,\hat{T}_q)$ has all the anchor nodes of all $q-p$ operations and the paths of backbone among these operations. $(\tilde{T}_i,\tilde{T}_j)$ has all the anchor nodes of all set $1$ operations and the paths of backbone among these operations. If all the grafting operations in the chain are independent, $(\hat{T}_p,\hat{T}_q)$ has the same substructure with $(\tilde{T}_i,\tilde{T}_j)$ after dropping the extra nodes. 
Inequality $CI(\hat{T}_p||\hat{T}_q)\geq (\tilde{T}_i||\tilde{T}_j)$ holds by Proposition 2 of \cite{binglin}.  $CI(T_i||T_j)\leq CI(T_p||T_q)$.

    Here we take a $T_1\leftrightarrow T_2\leftrightarrow T_3$ case as an example, as shown in Fig. \ref{f<}. We can simplify the calculation of $CI(T_1||T_3)$ and $CI(T_1||T_2)$ as shown in Fig. \ref{f<}. Then Proposition 2 of \cite{binglin} can tell us $CI(T_1||T_3)\geq CI(T_1||T_2)$. In the same way, we can conclude $CI(T_1||T_3)\geq CI(T_2||T_3)$.

\begin{figure}
  \centering
  \includegraphics[width=8cm]{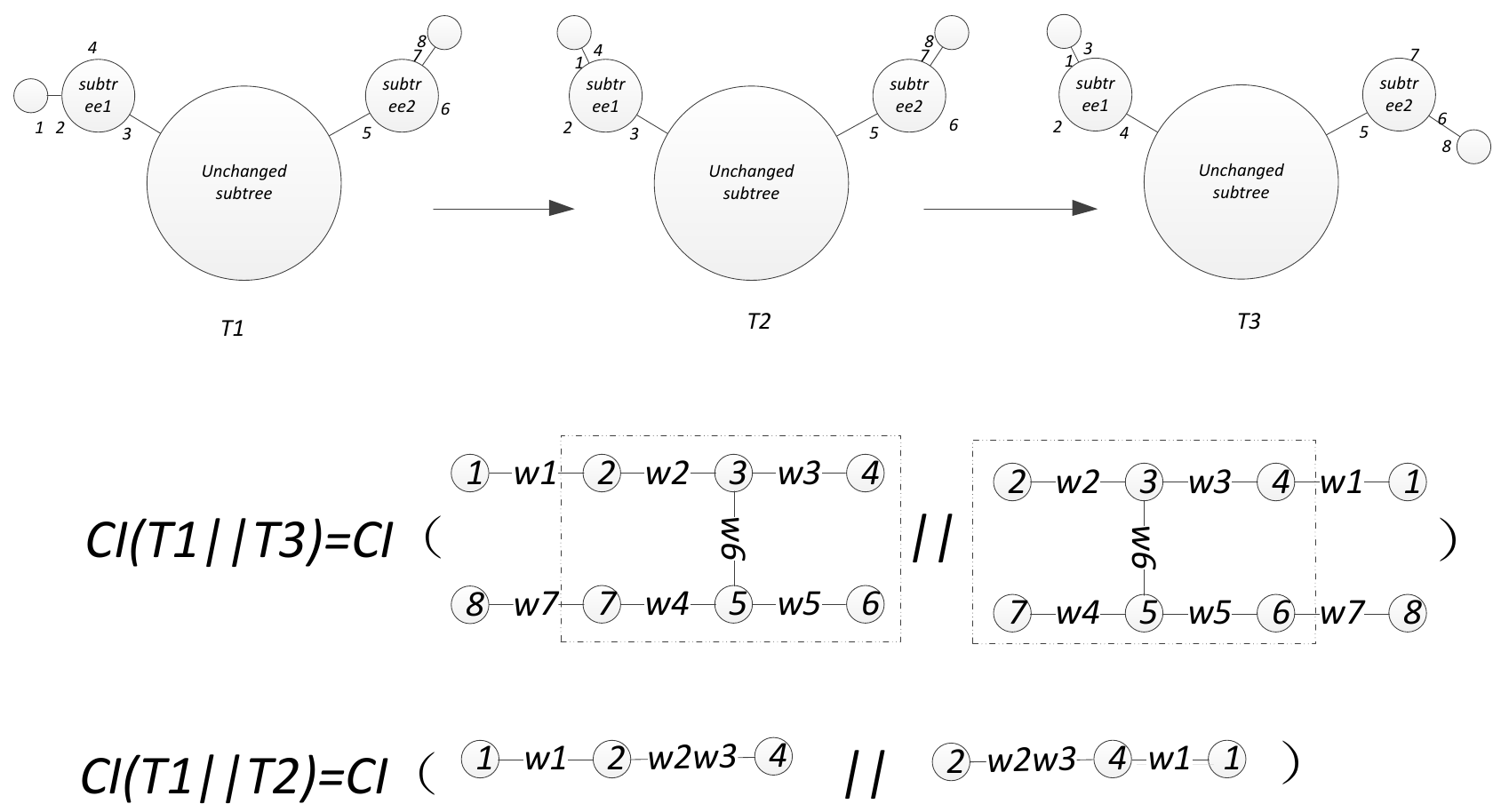}\\
  \caption{Example for Proposition \ref{<}}\label{f<}
\end{figure}

\section{Proof of Proposition~\ref{DR}}\label{A7}

{
\begin{align}
&D(\mathbf{\Sigma}_\lambda||\mathbf{\Sigma}_1)\nonumber\\
=&\frac{1}{2}\sum_i\left(\ln\left(\lambda+(1-\lambda)\mu_i\right)+\frac{1}{\lambda+(1-\lambda)\mu_i}-1\right)\\
&D(\mathbf{\Sigma}_\lambda||\mathbf{\Sigma}_2)\nonumber\\
=&\frac{1}{2}\sum_i\left(\ln\frac{\lambda+(1-\lambda)\mu_i}{\lambda_i}+\frac{\lambda_i}{\lambda+(1-\lambda)\mu_i}-1\right)\\
&CI(\mathbf{\Sigma}_1||\mathbf{\Sigma}_2)=\max_{\lambda\in[0,1]}\min\{D(\mathbf{\Sigma}_\lambda||\mathbf{\Sigma}_1),D(\mathbf{\Sigma}_\lambda||\mathbf{\Sigma}_2)\}\label{CI2}
\end{align}
Equation (\ref{CI2}) is another definition of Chernoff information\cite{cover2012elements}. The distances are only related to the eigenvalues we have chosen.}

We want to prove $CI(\hat{\mathbf{\Sigma}}_1||\hat{\mathbf{\Sigma}}_2)\geq CI(\tilde{\mathbf{\Sigma}}_1||\tilde{\mathbf{\Sigma}}_2)$ where $\hat{\mathbf{\Sigma}}_i=\mathbf{A}^*\mathbf{\Sigma}_i{\mathbf{A}^*}^T,
\tilde{\mathbf{\Sigma}}_i=\mathbf{D}\mathbf{\Sigma}_i\mathbf{D}^T$ for
arbitrary $N_O\times N$ matrix $\mathbf{D}$ and $i=1,2$.
Due to equation (\ref{CI2}), we only need to prove $D(\hat{\mathbf{\Sigma}}_\lambda||\hat{\mathbf{\Sigma}}_2)\geq D(\tilde{\mathbf{\Sigma}}_\lambda||\tilde{\mathbf{\Sigma}}_2)$ and $D(\hat{\mathbf{\Sigma}}_\lambda||\hat{\mathbf{\Sigma}}_1)\geq D(\tilde{\mathbf{\Sigma}}_\lambda||\tilde{\mathbf{\Sigma}}_1)$ where
$\hat{\mathbf{\Sigma}}_\lambda^{-1}=\lambda\hat{\mathbf{\Sigma}}_1^{-1}+(1-\lambda)\hat{\mathbf{\Sigma}}_2^{-1}$
and $\tilde{\mathbf{\Sigma}}_\lambda^{-1}=\lambda\tilde{\mathbf{\Sigma}}_1^{-1}+(1-\lambda)\tilde{\mathbf{\Sigma}}_2^{-1}$.

In the following part, we prove $D(\hat{\mathbf{\Sigma}}_\lambda||\hat{\mathbf{\Sigma}}_1)\geq D(\tilde{\mathbf{\Sigma}}_\lambda||\tilde{\mathbf{\Sigma}}_1)$ and the proof of $D(\hat{\mathbf{\Sigma}}_\lambda||\hat{\mathbf{\Sigma}}_2)\geq D(\tilde{\mathbf{\Sigma}}_\lambda||\tilde{\mathbf{\Sigma}}_2)$ is similar.
\begin{align}
D(\mathbf{\Sigma}_\lambda||\mathbf{\Sigma}_1)=\frac{1}{2}\sum_i g(\nu_i)
\end{align}
where $g(\nu_i)=\ln\left(\lambda+(1-\lambda)\nu_i\right)+\frac{1}{\lambda+(1-\lambda)\nu_i}-1$ and $\{\nu_i\}$ are generalized eigenvalues of $\mathbf{\Sigma}_1$ and $\mathbf{\Sigma}_2$.

$g(1)=0$. $g(\nu_i)$ is decreasing in $(0,1)$ and increasing in $(1,+\infty)$.
If $\nu_1^{(1)}\leq\nu_1^{(2)}\leq\nu_2^{(1)}\leq\nu_2^{(2)}\leq\dots\leq\nu_{N-1}^{(1)}\leq\nu_{N-1}^{(2)}\leq\nu_{N}^{(1)}$, we can choose $\mu_i^{(1)}=\nu_{i+1}^{(1)}$ for $\nu_i^{(2)}\geq 1$ and $\mu_i^{(1)}=\nu_{i}^{(1)}$ for $\nu_i^{(2)}< 1$ so that $\{\mu_i^{(1)}\}$ is an $(N-1)$ subset of $\{\nu_i^{(1)}\}$ and $\frac{1}{2}\sum_{i=1}^{N-1} g(\mu_i^{(1)})\geq\frac{1}{2}\sum_{i=1}^{N-1} g(\nu_i^{(2)})$.

If $m$ elements of $\{\nu_i^{(2)}\}$ are greater than one, $\{\mu_i^{(1)}\}$ contains the first $k$ elements and last $N-1-k$ elements of $\{\nu_i^{(1)}\}$.

We first introduce a preparation of proof.

\subsection{The relationship of eigenvalues}
We have four covariance matrices $\mathbf{\Sigma}_1,\mathbf{\Sigma}_2,\mathbf{\Sigma}_1'=\begin{bmatrix}\mathbf{\Sigma}_1&\mathbf{m}\\\mathbf{m}^T&a_0\end{bmatrix}$ and $\mathbf{\Sigma}_2'=\begin{bmatrix}\mathbf{\Sigma}_2&\mathbf{p}\\\mathbf{p}^T&b_0\end{bmatrix}$.

The eigenvalues of $\mathbf{\Sigma}_1\mathbf{\Sigma}_2^{-1}$ are $\mathbf{\lambda}_1^{(2)}\leq\mathbf{\lambda}_2^{(2)}\leq\dots\leq\lambda_{N-1}^{(2)}$.

The eigenvalues of $\mathbf{\Sigma}_1'\mathbf{\Sigma}_2'^{-1}$ are $\lambda_1^{(1)}\leq\lambda_2^{(1)}\leq\dots\leq\lambda_N^{(1)}$.

For $\mathbf{\Sigma}_1$ and $\mathbf{\Sigma}_2$, we can find an inverse matrix $\mathbf{P}$ where $\mathbf{\Sigma}_1^{(1)}=\mathbf{P}\mathbf{\Sigma}_1\mathbf{P}^T=Diag\{\lambda_1^{(2)},\lambda_2^{(2)},\dots,\lambda_{N-1}^{(2)}\}$ and $\mathbf{\Sigma}_2^{(1)}=\mathbf{P}\mathbf{\Sigma}_2\mathbf{P}^T=\mathbf{I}$.  $\mathbf{P}'=\begin{bmatrix} \mathbf{P}&\\&1\end{bmatrix}$ satisfies $\mathbf{\Sigma}_1'^{(1)}=\mathbf{P}'\mathbf{\Sigma}_1'\mathbf{P}'^T=\begin{bmatrix} \mathbf{\Sigma}_1^{(1)}&\mathbf{a}\\ \mathbf{a}^T&a_0\end{bmatrix}$ and $\mathbf{\Sigma}_2'^{(1)}=\mathbf{P}'\mathbf{\Sigma}_2'\mathbf{P}'^T=\begin{bmatrix} \mathbf{I}&\mathbf{b}\\ \mathbf{b}^T&b_0\end{bmatrix}$, where $\mathbf{a}=\{a_1,a_2,\dots,a_{N-1}\}^T$ and $\mathbf{b}=\{b_1,b_2,\dots,b_{N-1}\}^T$.

 $\{\lambda_k^{(1)}\}$ are the roots of $|\lambda\mathbf{\Sigma}_2'-\mathbf{\Sigma}_1'|=0$, which is equal to $F(\lambda)=0$ where
\begin{align}
F(\lambda)&=|\lambda\mathbf{\Sigma}_2'^{(1)}-\mathbf{\Sigma}_1'^{(1)}|=\left|\begin{matrix} \lambda \mathbf{I}-\mathbf{\Sigma}_1^{(1)}&\lambda\mathbf{b}-\mathbf{a}\\ \lambda\mathbf{b}^T-\mathbf{a}^T&\lambda b_0- a_0\end{matrix}\right|
\nonumber\\&=
\prod_{i=1}^{N-1}(\lambda-\lambda_i^{(2)})\times\nonumber\\&\left\{\lambda b_0- a_0-
(\lambda\mathbf{b}^T-\mathbf{a}^T){(\lambda \mathbf{I}-\mathbf{\Sigma}_1^{(1)})}^{-1}(\lambda\mathbf{b}-\mathbf{a})\right\}
\nonumber\\&=
\prod_{i=1}^{N-1}(\lambda-\lambda_i^{(2)})\times\left\{f_0(\lambda)-\sum_{i=1}^{N-1}\frac{{f_i^2(\lambda)}}{\lambda-\lambda_i^{(2)}}\right\}
\\&=|\mathbf{\Sigma}_2'^{(1)}|\lambda^N+\sum_{i=0}^{N-1} c_i\lambda^i
\end{align}
and $f_i(\lambda)=b_i\lambda-a_i$.

{At first, we assume that there is no multiple
eigenvalues for $\{\lambda_i^{(2)}\}$ and $f_i(\lambda_i^{(2)})\neq0$ for all $1\leq i\leq N-1$}.

  \begin{align}{(-1)}^{N-k}F(\lambda_{k}^{(2)})=&{(-1)}^{N-k+1}f_k^2(\lambda_{k}^{(2)})\prod_{i=1,i\neq k}^{N-1}(\lambda_{k}^{(2)}-\lambda_i^{(2)})\nonumber\\>&0
 \end{align}
We know $F(+\infty)=+\infty$, $F(-\infty)={(-1)^N}\infty$
 There is at least one root of $F(\lambda)=0$ in $(-\infty,\lambda_1^{(2)})$, $(\lambda_{N-1}^{(2)},+\infty)$ and $(\lambda_{i}^{(2)},\lambda_{i+1}^{(2)})$ for $1\leq i\leq N-2$.

We can conclude that $\lambda_1^{(1)}<\lambda_1^{(2)}<\lambda_2^{(1)}<\lambda_2^{(2)}<\dots<\lambda_{N-1}^{(1)}<\lambda_{N-1}^{(2)}<\lambda_{N}^{(1)}$.

{If $\lambda_i^{(2)}$ has multiplicity $n_i$, we can conclude that $\lambda_i^{(2)}$ has multiplicity $n_i-1$ in $F(\lambda)=0$. If we put this exact multiple eigenvalues aside, other eigenvalues satisfy the ordering before.}

{If $f_i(\lambda_i^{(2)})=0$, we can conclude that $\lambda_i^{(2)}$ is also the root of $F(\lambda)=0$.  If we put this eigenvalue aside, other eigenvalues satisfy the ordering before.}

For all cases, $\lambda_1^{(1)}\leq\lambda_1^{(2)}\leq\lambda_2^{(1)}\leq\lambda_2^{(2)}\leq\dots\leq\lambda_{N-1}^{(1)}\leq\lambda_{N-1}^{(2)}\leq\lambda_{N}^{(1)}$.

\subsection{Main proof}

For $N$-dimension graphs whose covariance matrices are $\mathbf{A}$ and $\mathbf{B}$, we can do inverse linear transformation by $\mathbf{A}'=\mathbf{K}\mathbf{A}\mathbf{K}^T$ and $\mathbf{B}'=\mathbf{K}\mathbf{B}\mathbf{K}^T$. Then we can do dimension-reduction from $N$ to $N-1$ by $\mathbf{A}''=[\mathbf{I}_{N-1},0]\mathbf{A}'[\mathbf{I}_{N-1},0]^T$ and $\mathbf{B}''=[\mathbf{I}_{N-1},0]\mathbf{B}'[\mathbf{I}_{N-1},0]^T$.

The generalized eigenvalues of $\mathbf{A}$ and $\mathbf{B}$ are $\lambda_1^{(1)}\leq\lambda_2^{(1)}\leq\dots\leq\lambda_N^{(1)}$, which are also generalized eigenvalues of $\mathbf{A}'$ and $\mathbf{B}'$.
The generalized eigenvalues of $\mathbf{A}''$ and $\mathbf{B}''$ are $\lambda_1^{(2)}\leq\lambda_2^{(2)}\leq\dots\leq\lambda_{N-1}^{(2)}$. We can conclude $\lambda_1^{(1)}\leq\lambda_1^{(2)}\leq\lambda_2^{(1)}\leq\lambda_2^{(2)}
\leq\dots\leq\lambda_{N-1}^{(1)}\leq\lambda_{N-1}^{(2)}\leq\lambda_{N}^{(1)}$.

We can also do the procedure to reduce the dimension from $N-1$ to $N-2$ and get $\lambda_1^{(2)}\leq\lambda_1^{(3)}\leq\lambda_2^{(2)}\leq\lambda_2^{(3)}
\leq\dots\leq\lambda_{N-2}^{(2)}\leq\lambda_{N-2}^{(3)}\leq\lambda_{N-1}^{(2)}$.

The generalized eigenvalues after $k$ dimension reduction are $\{\lambda_i^{(k+1)}\}$ of size $N-k$.

In No. $(N-N_O)$ dimension reduction stage, we can always find an $N_O$ subset $\{\mu_i^{(N-N_O)}\}$ of $\{\lambda_i^{(N-N_O)}\}$ so that $\frac{1}{2}\sum_{i=1}^{N_O} g\left( \mu_i^{(N-N_O)}\right)\geq \frac{1}{2}\sum_{i=1}^{N_O}g \left( \lambda_i^{(N-N_O+1)}\right)$ because $\lambda_1^{(N-N_O)}\leq\lambda_1^{(N-N_O+1)}\leq\lambda_2^{(N-N_O)}\leq\lambda_2^{(N-N_O+1)}
\leq\dots\leq\lambda_{N_O+1}^{(N-N_O)}\leq\lambda_{N_O}^{(N-N_O+1)}\leq\lambda_{N_O+1}^{(N-N_O)}$.

We can always find $N_O$ subset $\{\mu_i^{(N-N_O-1)}\}$ of $\{\lambda_i^{(N-N_O-1)}\}$ and $\{\mu_i^{(N-N_O)}\}$ of $\{\lambda_i^{(N-N_O)}\}$, which satisfy $\frac{1}{2}\sum_{i=1}^{N_O}g \left( \mu_i^{(N-N_O-1)}\right)\geq \frac{1}{2}\sum_{i=1}^{N_O}g \left( \mu_i^{(N-N_O)}\right)\geq\frac{1}{2}\sum_{i=1}^{N_O} g\left( \lambda_i^{(N-N_O+1)}\right)$ with the same reason.

With the same method, we can conclude that there exists an  $N_O$ subset $\{\mu_i^{(1)}\}$ of $\{\lambda_i^{(1)}\}$ so that $\frac{1}{2}\sum_{i=1}^{N_O} g\left( \mu_i^{(1)}\right)\geq\frac{1}{2}\sum_{i=1}^{N_O} g\left( \lambda_i^{(N-N_O+1)}\right)$. If $k$ elements of $\{\lambda_i^{(N-N_O+1)}\}$ are greater than one, $\{\mu_i^{(1)}\}$ contains the first $k$ elements and last $N_O-k$ elements of $\{\lambda_i^{(1)}\}$.

$\frac{1}{2}\sum_{i=1}^{N_O} g\left( \mu_i^{(1)}\right)$ is $D({\mathbf{\Sigma}}_\lambda||{\mathbf{\Sigma}}_1)$  with linear transformation matrix $\mathbf{A}_k$. $\frac{1}{2}\sum_{i=1}^{N_O} g\left( \lambda_i^{(N-N_O+1)}\right)$ is $D(\tilde{\mathbf{\Sigma}}_\lambda||\tilde{\mathbf{\Sigma}}_1)$ under arbitrary linear transformation where the arbitrariness is embodied in $\mathbf{K}$.

Inequality $D({\mathbf{\Sigma}}_\lambda||{\mathbf{\Sigma}}_1)\geq D(\tilde{\mathbf{\Sigma}}_\lambda||\tilde{\mathbf{\Sigma}}_1)$ holds for arbitrary $\mathbf{D}$ and its corresponding $\mathbf{A}_k$.

In the same way, $D({\mathbf{\Sigma}}_\lambda||{\mathbf{\Sigma}}_2)\geq D(\tilde{\mathbf{\Sigma}}_\lambda||\tilde{\mathbf{\Sigma}}_2)$ holds for arbitrary $\mathbf{D}$ and its corresponding $\mathbf{A}_k$.

Inequality $CI(\hat{\mathbf{\Sigma}}_1||\hat{\mathbf{\Sigma}}_2)\geq CI(\tilde{\mathbf{\Sigma}}_1||\tilde{\mathbf{\Sigma}}_2)$ for arbitrary $\mathbf{D}$ and $\mathbf{A}^*$ is the optimal matrix in the set $\{\mathbf{A}_k|N_O+m-N\leq k\leq m, k\geq0\}$.

\end{document}